\documentclass[prb,aps,twocolumn,superscriptaddress]{revtex4-2} 
\usepackage{graphicx}
\usepackage{dcolumn}
\usepackage{bm}
\usepackage{lipsum}
\usepackage{physics}
\usepackage{ragged2e}
\usepackage{soul}
\usepackage{braket}
\usepackage{bbm}
\usepackage{lipsum}
\usepackage{color}
\usepackage{xcolor}
\usepackage{amsfonts}
\usepackage{amsmath} 
\usepackage{amsmath,amssymb,latexsym}
\usepackage{wasysym,stmaryrd,marvosym}
\usepackage{mathrsfs}
\usepackage{dsfont}
\usepackage{float}
\providecommand{\keywords}[1]{\textbf{\textit{Index terms---}} #1}
\usepackage[mathlines]{lineno}
\usepackage[colorinlistoftodos]{todonotes}
\usepackage[colorlinks=true, allcolors=blue]{hyperref}
\usepackage{array}

\newcommand{\Ham}{\mathcal{H}}
\newcommand{\F}{\mathcal{F}}
\newcommand{\E}{\mathcal{E}}
\newcommand{\Keka}{\alpha-\mathcal{T}_3}
\newcommand{\J}{\mathcal{J}}

\newcommand{\re}[1]{ \mathbbm{R} \textup{e}  \{ #1 \} }
\newcommand{\im}[1]{ \mathbbm{I} \textup{m}  \{ #1 \} }

\definecolor{darkred}{rgb}{0.8,0.1,0.1}
\begin{document}
\newcommand{\uabc}{Facultad de Ciencias, Universidad Aut\'onoma de Baja California, Apartado Postal 1880, 22800 Ensenada, Baja California, M\'exico}
\newcommand{\ohio}{Department of Physics and Astronomy and Nanoscale and Quantum Phenomena Institute, Ohio University, Athens, Ohio 45701, USA}
\newcommand{\cnyn}{Centro de Nanociencias y Nanotecnolog\'ia, Universidad Nacional Aut\'onoma de M\'exico, Apartado Postal 2681, 22800 Ensenada, Baja California, M\'exico}

\title{Band structure and optical response of Kekulé-modulated $\alpha-\mathcal{T}_3$ model}

\author{Luis E. S\'anchez-Gonz\'alez}
\affiliation{\uabc}
\author{M. A. Mojarro}
\affiliation{\ohio}
\author{Jes\'us A. Maytorena}
\affiliation{\cnyn}
\author{R. Carrillo-Bastos}
\email{ramoncarrillo@uabc.edu.mx}
\affiliation{\uabc}

\date{\today}

\begin{abstract}
We study the electronic band structure and optical response of a hybrid model, a $\Keka$ model featuring a $\sqrt{3}\times\sqrt{3}$ Kekulé pattern modulation. Such a hybrid system may result from the depositing of adatoms in a hexagonal lattice, where the two sublattices are displaced in the perpendicular direction, like in germanene and silicene. We derive analytical expressions for the energy dispersion and the eigenfunctions using a tight-binding approximation of nearest-neighbor hopping electrons.
The energy spectrum consists of a double-cone structure with Dirac points at zero momentum caused by Brillouin zone folding and a doubly degenerate flat band resulting from destructive quantum interference effects. 
Furthermore, we study the spectrum of intraband and interband transitions through the joint density of states, the optical conductivity, and the Drude spectral weight. 
We find new conductivity terms resulting from the opening of intervalley channels that are absent in the $\Keka$  model and manifest themselves as van Hove singularities in the optical response. 
In particular, we identify an absorption window related to intervalley transport, which serves as a viable signature for detecting Kekulé periodicity in two-dimensional materials. 
\end{abstract}

 
\maketitle

\section{Introduction} 

In recent years, two-dimensional (2D) materials exhibiting flat bands have garnered significant attention due to their unique electronic and transport properties, making them ideal platforms for exploring novel physical phenomena~\cite{KopninSCTopologicalFlatBand2011, cao2018unconventional, EhlenOriginFlatBand2020, Leykam2018ArtificialExperiments, drost2017topologicalstates, AlEzziTopologicalFlatBand2024, BaoFlatBand2022, EscuderoDesignigMoireStrain2024, de2024flat, deng2003origin, TopologicalFlatBandTaboada2017}. The observation of correlated insulator states and signatures of unconventional superconductivity in twisted bilayer graphene~\cite{cao2018unconventional} has further fueled the interest in systems hosting flat bands close to the Fermi level ~\cite{MacDonaldMoireBands2011, mogera2020new}.  

Line graphs \cite{line1-cvetkovic2004spectral,line2-PhysRevResearch.4.023063,Leykam2018ArtificialExperiments,Kollar2020}, such as the kagome and pyrochlore lattices, along with bipartite lattices like the Lieb and dice lattices \cite{Lieb1989,Leykam2018ArtificialExperiments}, naturally host flat bands in their energy spectrum thanks to destructive interference between electron wavefunctions. 
The $\Keka$ model~\cite{Sutherland1986LocalizationTopology, Bercioux2009MasslessLattice, Raoux2014FromFermions,Mojarro20202} is a simple example of flat-band system which continuously evolves between the graphene and dice lattice by modulation of a hopping parameter.
Its crystal structure consists of a honeycomb lattice (rim atom), with an additional site at the center of each hexagon (a hub atom) that couples to neighboring atoms with only one of the sublattices, hosting a flat band and Dirac cones close to the Fermi level.
Numerous studies are dedicated to unraveling the mechanisms behind the emergence of flat bands in Dirac systems~\cite{deng2003origin, Oriekhov2018ElectronicRibbon, Tarnopolsky2019OriginGraphene} and how they give rise to a variety of quantum phases~\cite{Yu2018ChernLattices, Yuan2018ModelBeyond}. The optical response of flat bands has also been studied~\cite{Illes2015HallModel, HanOpticalResponseFlatBand2022, iurovoptical2023, ye2024optical}, but since the group velocity in these bands vanishes, identifying clear optical signatures in the low-frequency range seems challenging.

On the other hand, spatial bond modulation can induce exotic effects in
the electronic properties of two-dimensional materials~\cite{ponomarenko2013cloning, houelectron2007, yankowitz2012emergence, EscuderoDesignigMoireStrain2024, park2008anisotropic}. One of the most interesting examples of spatial modulation is the Kekulé distortion in the graphene honeycomb lattice~\cite{Bao2021ExperimentalGraphene, Gomes2012DesignerGraphene, Gutierrez2016ImagingGraphene, Gamayun2018Valley-momentumTexture}, 
where the lattice acquires a bond density wave with superlattice unit cell $\sqrt{3}\times\sqrt{3}$ larger than the original unit cell. As a result, Brillouin zone folding brings the $K, K '$ points to the center of the Brillouin zone ($\Gamma$ point). Experiments suggest two types of Kekulé modulations in graphene with distinct low-energy spectrum ~\cite{Gamayun2018Valley-momentumTexture, eomkekule2020}: the so-called Kek-Y phase with two Dirac cones with different velocities, and the Kek-O phase with a doubly degenerate massive Dirac band. Several mechanisms have been proposed to generate phases with different Kekulé distortions~\cite{quubiqutuos2022, Cheianov2009HiddenGraphene, Cheianov2009OrderedGraphene, Farjam2009EnergyCalculations, Sugawara2011FabricationGraphene, Kanetani2012CaCa, Chamon2000SolitonsNanotubes, Classen2014InstabilitiesInteractions, Weeks2010Interaction-drivenSemimetal, Hou2007ElectronStructures, Marianetti2010FailureTension, Lee2011BandGraphene, Giovannetti2015KekuleBilayer, Im2023ModifiedHeterostructure, Ye2023KekuleSuperlattices}, indicating its ubiquitousness in hexagonal lattices~\cite{quubiqutuos2022}. Electrical and optical signatures offer a promising avenue for studying and understanding the mechanism behind the Kekulé phase~\cite{HerreraElectronicOpticalKekY, HerreraDynamicPlasmonsKekule2020, ValleydrivenSantacruz2022, AndradeKekuléValleyBire2022, HerreraOptoelectronic2021, MohammadiOpticalKek2021, MohammadiElectronicKek2022,Mojarro2020}.

From a topological perspective, Kekulé-distorted graphene was first proposed as a novel platform hosting fractionally charged topological
excitations~\cite{houelectron2007}. Mechanical strain applied to graphene-based heterostructures with Kekulé patterning also gives rise to intriguing topological effects~\cite{tajkov2020}. Moreover, Kekulé distortion is one
of the suggested mechanisms behind the superconducting and correlated insulating states
behavior in magic-angle twisted bilayer graphene ~\cite{XuKekule2018, RoyKekule2010, Po2018, Da2019, Huangslave2020, localkekule2022}, further increasing the interest in the study of Kekulé-patterned superlattices.

\begin{figure*}
\centering
\includegraphics[scale=0.295]{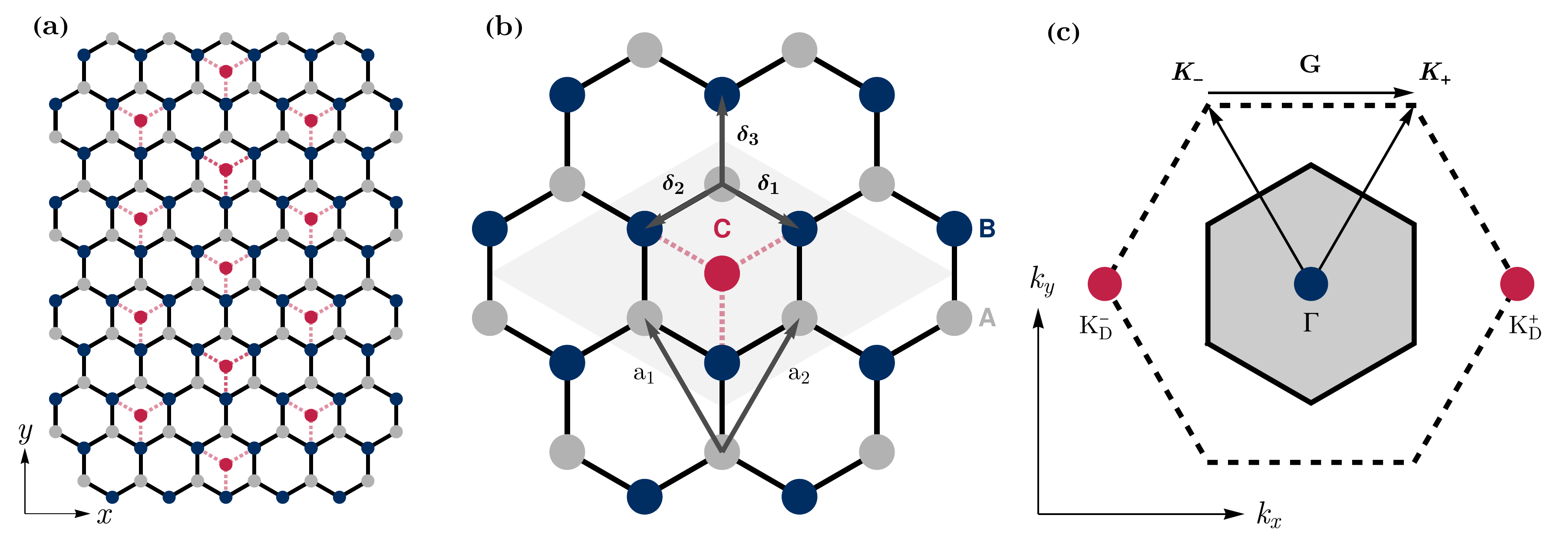}
\caption{(a) Lattice structure of Kekulé-modulated $\Keka$ model. Atoms on sublattices A and B are depicted as gray and blue circles, respectively. The hub atoms in red appear with Kekulé periodicity and only connect with the B sublattice. (b) The basis vectors of the honeycomb lattice are $\vb{a}_1$ and $\vb{a}_2$. The vectors $\boldsymbol{\delta}_1$, $\boldsymbol{\delta}_2$, and $\boldsymbol{\delta}_3$ connect each A or B atom to its nearest neighbors. The gray parallelogram represents the unit cell of the superlattice.  (c) Reciprocal space with vectors $\vb{K}_{\pm}$. The original (honeycomb lattice) Brillouin zone is represented as a black dashed hexagon. The $K_{D} ^{\pm}$ valleys (at the red Dirac points) are coupled by the wave vector $\vb{G} = \vb{K}_{+} - \vb{K}_{-}$ and folded onto the center of the superlattice Brillouin zone (blue point).   } \label{fig:fig1}
\end{figure*}

In this work, we propose a hybrid model based on a honeycomb lattice with an atom located at the center of each hexagon, appearing only with Kekulé periodicity. We name this system as ``Kekulé-modulated $\Keka$ model" (Kek-$\alpha$), which provides a robust platform for studying valley and flat-band physics. The feasibility of constructing such a model has been discussed in the literature~\cite{Bao2021ExperimentalGraphene, EhlenOriginFlatBand2020, quubiqutuos2022, Wang2DSupeconductivity2014, modulationarraga2018, kekuletexturesgianluca2015}. For instance, a hexagonal lattice where its sublattices are displaced along the $z$-plane, similar to silicene or related systems~\cite{GarciaJoelsonGroupIVGraphene}, with atoms deposited with Kekulé periodicity, could be well described by our model. Given the increasing interest in space-modulated and flat-band materials, we aim to anticipate potential future developments in these systems.

This paper is organized as follows. In Sec.~\ref{sec:secII} we present the tight-binding model and in Sec.~\ref{sec:secIII} we derive a Dirac-like Hamiltonian and its energy dispersion. 
In Sec.~\ref{sec:secIV} we study the optical transitions. We first study the joint density of states to identify critical frequencies, which will determine the prominent spectral features of the optical response (Sec.~\ref{subsec:secIIIA}). The optical conductivity, due to intra and interband transitions, is calculated within the Kubo formalism in Sec.~\ref{subsec:secIIIB}. Finally, we present our conclusions and remarks in Sec.~\ref{sec:secV}.

\section{Tight-binding model} \label{sec:secII}

We consider a honeycomb lattice (like graphene, germanene, or silicene) with adatoms on its surface disposed with a $\sqrt{3}\times\sqrt{3}$  Kekulé periodicity. 
We call this superlattice with seven atoms per unit cell, Kekulé-modulated $\Keka$ model.
The adatoms are located on top of the center of the hexagon (hub atoms), forming a third triangular sublattice C with a larger lattice parameter that only couples with sublattice B, as shown in Fig.~\hyperref[fig:fig1]{1(a)}. 
The corresponding 
tight-binding Hamiltonian is:
\begin{equation} \label{eq:Hamiltonian}
H = - t \displaystyle\sum_{\vb{r}} \sum_{j = 1} ^{3}  b^{\dagger}_{\vb{r}} a_{\vb{r} - \boldsymbol{\delta}_j} -  \sum_{\vb{r}'} \sum_{j =1} ^{3} t'   c_{\vb{r}'} ^{\dagger} b_{\vb{r}' - \boldsymbol{\delta}_j} + \textup{H.c.},
\end{equation} 
where the first term describes the honeycomb lattice, with $t$ the hopping energy between nearest neighbor sites belonging to sublattices A and B, connected by the vectors 
$\boldsymbol{\delta}_1 = \frac{a}{2} (\sqrt{3}, -1), $ 
$\boldsymbol{\delta}_2 = - \frac{a}{2} (\sqrt{3}, 1), $ 
$\boldsymbol{\delta}_3 = a(0,1)$.  
Here $a$ is the atomic distance. Thus, the honeycomb lattice vectors are 
$\vb{a}_1 = \boldsymbol{\delta}_3 - \boldsymbol{\delta}_1$ and $\vb{a}_2 = \boldsymbol{\delta}_3 - \boldsymbol{\delta}_2$, such that the rim atoms have positions $\vb{r} = n \vb{a_1} + m \vb{a_2} \; (n,m \in \mathbb{Z})$ in sublattice B and $\vb{r} + \boldsymbol{\delta}_3$ in sublattice A (see Fig.~\hyperref[fig:fig1]{1(b)}).  
The second term describes 
the coupling between hub atoms  at $\vb{r}' = n (2 \vb{a}_1 - \vb{a}_2) + m (2\vb{a}_2 - \vb{a}_1)$
and nearest neighbor atoms in sublattice B, with $t'=\alpha t$ the corresponding hopping energy. 
The parameter $\alpha$ varies continuously between 0 and 1, interpolating between the honeycomb ($\alpha=0$) and a ``partial" dice lattice ($\alpha=1$).


In order to describe the superlattice (Kek-$\alpha$) with the hexagonal Brillouin zone, we consider additional sites added in sublattice C but with zero amplitude hoppings, in such a way that we can sum over all the cells $\vb{r}' = \vb{r} +\vb{\delta_3}$ replacing $t' \rightarrow t_{\vb{r},j}$, with
\begin{align}
t_{\vb{r}, j}  & = \alpha t \{ 1 + 2 \mathbb{R} e [ e^{i \vb{G} \cdot (\vb{r} + \boldsymbol{\delta}_j ) } ]\}, 
\end{align}
where the Kekulé wave vector is the ``distance'' between valleys, defined as $\vb{G} = \vb{K}_{+} - \vb{K}_{-}$ (see Fig.~\hyperref[fig:fig1]{1(b)}). 
\begin{figure}
    \centering
    \includegraphics[scale=0.8]{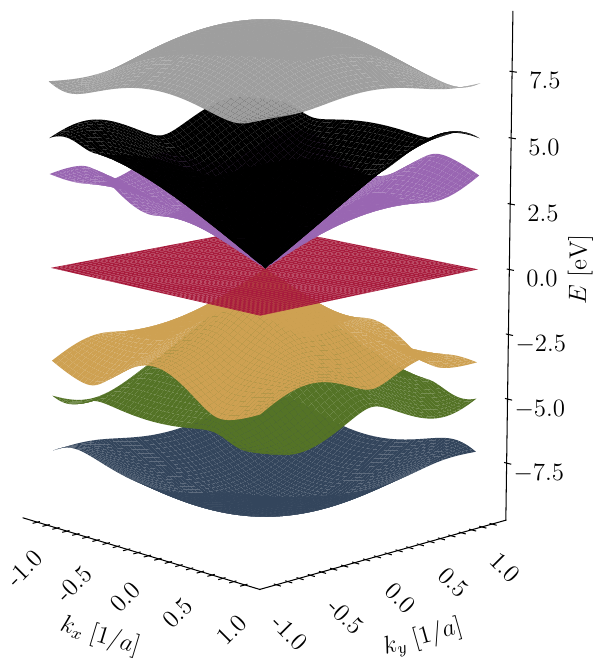}
    \caption{Energy band structure of the Kek-$\alpha$ model obtained by a direct diagonalization of the tight-binding Hamiltonian defined by the lattice shown in Fig.~\hyperref[fig:fig1]{1(b)}. We consider $t=2.7\;\textup{eV}$ and $\alpha=1$. The spectrum displays dispersive bands plus a flat band at zero energy originating from the bonded hub atoms. }
    \label{fig:fig2}
\end{figure}
The corresponding Hamiltonian in momentum space is given by
\begin{align}
H  = & - \sum_{\vb{k}} a_{\vb{k}} ^{\dagger} f(\vb{k}) b_{\vb{k}} -\alpha \sum_{\vb{k}} \left[ b_{\vb{k}} ^{\dagger} f (\vb{k}) c_{\vb{k}} + \right.  \nonumber \\
&  \left.  +\, b_{\vb{k}+ \vb{G}} ^{\dagger} f (\vb{k} + \vb{G}) c_{\vb{k} } + b_{\vb{k} - \vb{G}} ^{\dagger} f (\vb{k} - \vb{G} ) c_{\vb{k}} \right] + \textup{H.c.}\,,
\end{align}
where we have defined,
\begin{equation}
    f(\vb{k}) = t \sum_{j=1} ^{3} e^{i \vb{k} \cdot \boldsymbol{\delta}_j } .
\end{equation}
with the momentum $\vb{k}$ varying in the original (honeycomb lattice) Brillouin zone. In order to restrict $\vb{k}$ to the superlattice Brillouin zone, we group the annihilation operators at $\vb{k}$ and $\vb{k} \pm \vb{G}$ in the column vector $\Psi_{\vb{k}} = (a_{\vb{k}} , a_{\vb{k} - \vb{G}} , a_{\vb{k} + \vb{G}} , b_{\vb{k}} , b_{\vb{k} - \vb{G}} , b_{\vb{k} + \vb{G}}, c_{\vb{k}}, c_{\vb{k} - \vb{G}}, c_{\vb{k} + \vb{G}})^{\text{T}}$, and write the Hamiltonian in a $9\times 9$ matrix form:
\begin{equation}
H = - \Psi_{\vb{k}} ^{\dagger} \Ham (\vb{k}) \Psi_{\vb{k}},
\end{equation}
\noindent where

\begin{equation} \label{eq:Hamkk}
\Ham (\vb{k}) = \begin{pmatrix}
\vb{0} & \mathcal{F} (\vb{k}) & \vb{0} \\
\mathcal{F} ^{\dagger} (\vb{k}) & \vb{0} & \alpha \mathcal{E} (\vb{k}) \\
\vb{0} & \alpha \mathcal{E} ^{\dagger} (\vb{k}) & \vb{0}
\end{pmatrix},
\end{equation}
and
\begin{equation}
\mathcal{F} (\vb{k}) = \begin{pmatrix}
f_0 & 0 & 0 \\
0 & f_{-1} & 0 \\
0 & 0 & f_{1}
\end{pmatrix}, \;\; \mathcal{E} (\vb{k}) = \begin{pmatrix}
f_{0} &  f_{0} &  f_{0} \\
 f_{-1} & f_{-1} &  f_{-1} \\
 f_{1}&  f_{1} & f_{1}
\end{pmatrix},
\end{equation}
with $f_n (\vb{k} + n\vb{G})$, and we have used the relation $f (\vb{k} \pm 2 \vb{G}) = f (\vb{k} \mp \vb{G}) $~\cite{Gamayun2018Valley-momentumTexture}.

In Fig.~\ref{fig:fig2}, we show the energy dispersion of Kek-$\alpha$ superlattice. The Kekulé periodicity brings the valleys into the $\Gamma$ point, as observed in Kekulé-distorted graphene~\cite{Gamayun2018Valley-momentumTexture}, with the addition of a flat band due to the presence of hub sites. 
In the following section, we will derive an effective  Hamiltonian for this hybrid system in the low energy limit.

\section{Low energy Hamiltonian} \label{sec:secIII}

An effective Hamiltonian for low energies can be obtained considering $\alpha \ll 1$ and by noticing that the rows and columns of the matrices $\F$ and $\E$ associated with modes $a_{\vb{k}}, b_{\vb{k}},$ and $c_{\vb{k}}$ (illustrated in blue and gray in Fig.~\ref{fig:fig2}) lead to high energy bands, thus negligible in the low energy limit. Consequently, in this limit the spectrum is primarily determined by six modes, denoted as $u_{\vb{k}} = ( a_{\vb{k} - \vb{G}}, a_{\vb{k} + \vb{G}}, b_{\vb{k} - \vb{G}}, b_{\vb{k} + \vb{G}}, c_{\vb{k} - \vb{G}}, c_{\vb{k} + \vb{G}})$. Projecting onto this subspace results in the reduction of the nine-band Hamiltonian to an effective six-band Hamiltonian
\begin{equation} \label{eq:Hameff}
    H_{\textup{eff}} = u^{\dagger} _{\vb{k}} \begin{pmatrix}
        \vb{0} & g (\vb{k}) & \vb{0} \\
g ^{\dagger} (\vb{k}) & \vb{0} & \alpha h (\vb{k}) \\
\vb{0} & \alpha h ^{\dagger} (\vb{k}) & \vb{0}
    \end{pmatrix} u_{\vb{k}},
\end{equation}
where
\begin{equation}
    g (\vb{k}) = \begin{pmatrix}
    f_{-1} & 0 \\
    0 & f_{1}
    \end{pmatrix}, \qquad
    h (\vb{k})  = \begin{pmatrix}
        f_{-1} & f_{-1 } \\
        f_{1 } & f_{1}
    \end{pmatrix}.
\end{equation}
We identify the $K$ valley with $+\vb{G}$ and the $K'$ valley with $- \vb{G}$. 
The $\vb{k}$-dependence of $f_{\pm 1}$ may be linearized near $\vb{k} = 0$, leading to $f_{\pm 1}(k_x, k_y) = \hslash v_F (\mp k_x + i k_y ) $, where $v_F = 3 at/2 \hslash$ is the Fermi velocity.  Finally, we can write a Dirac-like equation for the  Kek-$\alpha$ model as
\begin{equation} \label{eq:DiracEq}
    \Ham \begin{pmatrix}
        \Psi_{K} \\ \Psi_{K'}
    \end{pmatrix} = \varepsilon
    \begin{pmatrix}
        \Psi_{K} \\ \Psi_{K'}
    \end{pmatrix}, \;  \Ham  = \begin{pmatrix}
        \hbar v_F \vb{k} \cdot \vb{S} & Q \\
        Q^{\dagger} & \hbar v_F \vb{k} \cdot \vb{S}^*
    \end{pmatrix},
\end{equation}
\begin{equation}
    \Psi_{K} = \begin{pmatrix}
        \psi_{A, K} \\ -\psi_{B, K} \\ \psi_{C, K} 
    \end{pmatrix}, \qquad \Psi_{K'} = \begin{pmatrix}
        \psi_{A, K'} \\ \psi_{B, K'} \\ \psi_{C, K'} 
    \end{pmatrix},
\end{equation}

\begin{equation}
    Q =  \alpha \hbar v_F  \begin{pmatrix}
        0 & 0 & 0 \\
        0 & 0 & k e^{i\theta_{\boldsymbol{k}}} \\
        0 &   k e^{i\theta_{\boldsymbol{k}}} & 0 
    \end{pmatrix},
\end{equation}
where $k =  \vert \vb{k} \vert$, $\theta_{\vb{k}} = \tan ^{-1} (k_x/k_y)$,
$\vb{S}  = (S_x, S_y)$. The pseudospin operators $S_x$ and $S_y$ are defined as
\begin{align}
    S_x  = \begin{pmatrix}
        0 & 1  & 0 \\
        1 & 0 & \alpha \\
        0 & \alpha & 0 
    \end{pmatrix}, 
     \ \ \,  S_y  = \begin{pmatrix}
        \;0\;\; & -i  & 0 \\
        \;i\;\; & 0 & -\alpha i \\
        \;0\;\; & \alpha i & 0 
    \end{pmatrix}.
\end{align}
Note that when $\alpha = 0$, the set $\{S_x, S_y\}$ corresponds to an effective spin-1/2 algebra. 
In the case where $\alpha = 1$, it forms a spin-1 algebra. Therefore, similar to the $\Keka$ model, this can be interpreted as a smooth interpolation between $S = 1/2$ and $S = 1$ structures. 

\begin{figure}
    \centering
    \includegraphics[scale=0.34]{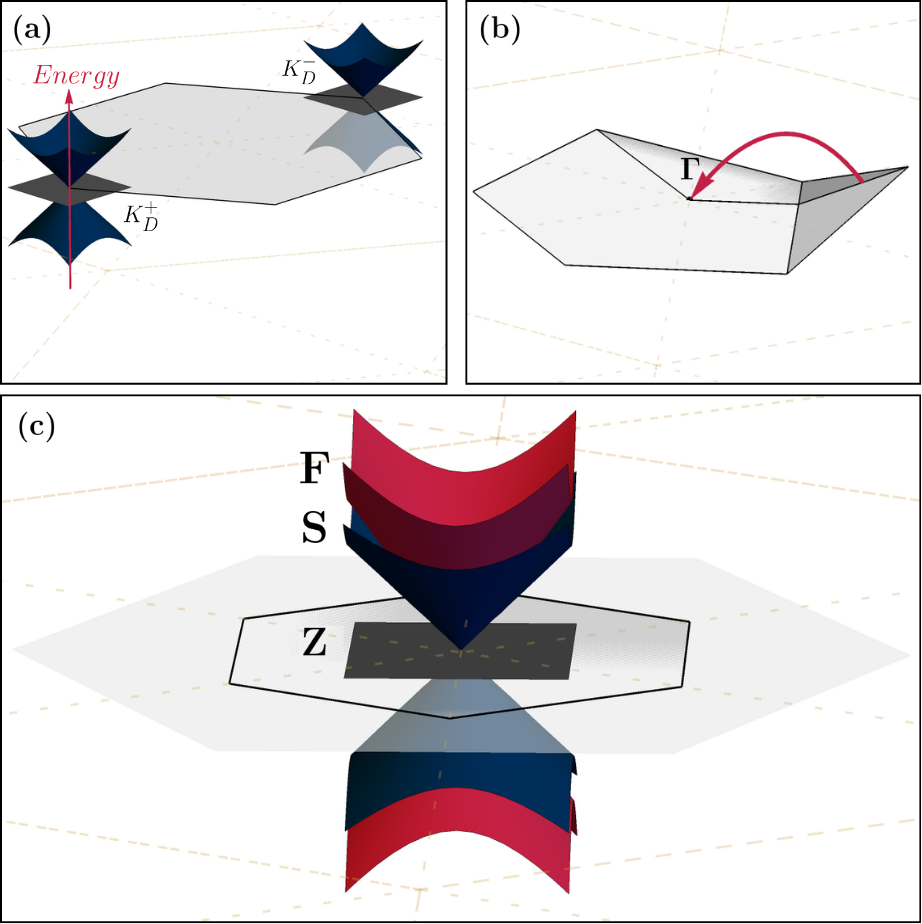}
    \caption{(a) Brillouin zone of the $\Keka$ model in reciprocal space is shown as a gray hexagon. The two inequivalent valleys are situated in the corners of the hexagon. (b) Brillouin zone folding due to the Kekulé periodicity. (c) Energy dispersion relation of Kek-$\alpha$ around the $\Gamma$ in the folded Brillouin zone. The label $\textbf{Z}$ designates the flat band with ``zero'' velocity, $\textbf{S}$ denotes the ``slow'' cone with velocity $v_F$, and $\textbf{F}$ signifies the ``fast'' cone with velocity $\Delta_\alpha v_F$. }
    \label{fig:fig3}
\end{figure}

The low-energy band structure of Kek-$\alpha$ exhibits a distinctive double cone structure, referred to as ``fast'' (\textbf{F}$_\pm$) and ``slow'' (\textbf{S}$_\pm$) cones, along with a doubly-degenerate flat band denoted as ``zero'' (\textbf{Z}). The energy dispersion relation is given by
\begin{equation} \label{eq:energy}
    \varepsilon_{\eta} ^{\xi} (\vb{k}) = \eta \hslash v_F (\Delta_{\alpha} )^{\xi} k
\end{equation}
where $\Delta_{\alpha} = \sqrt{4\alpha^2+ 1}$, $\eta$ is the index band, $\eta=1$ for the conduction band, $\eta=-1$ for the valence band, and $\eta=0$ for the flat band. 
The index $\xi = \{0,1\}$ labels the two degenerate flat bands ($\textbf{Z}^\xi$) and the two cones, defining two velocities: the `fast velocity' $v_F\Delta_{\alpha}$ ($\xi=1$) and the `slow velocity' $v_F$ ($\xi=0$), associated to the fast cones ${\bf F}_{\pm}$ and slow cones ${\bf S}_{\pm}$, respectively.

%

In the $\Keka$ model, rescaling the energy renders the spectrum independent of $\alpha$~\cite{Raoux2014FromFermions, Illes2015HallModel}. However, in our case, such rescaling is not possible, and the spectrum remains $\alpha$-dependent because the Kekulé term couples with one sublattice only. Thus, when the two valleys fold onto the $\Gamma$ point (see Fig.~\hyperref[fig:fig3]{3(b)}) one of the cones shows a strong dependence on $\alpha$ and the other remains independent.

The eigenfunctions $\psi_{\eta} ^{\xi}$ are given by

\begin{equation}
    \psi_{0} ^{0} (\vb{k}) = \frac{1}{\sqrt{2}} \begin{pmatrix}
        0 \\ 0 \\ -1 \\ 0 \\ 0 \\ 1
    \end{pmatrix}, \;\;\;\;
    \psi_{0} ^{1} (\vb{k}) = \frac{1}{\sqrt{8\alpha^2 + 2}} \begin{pmatrix} 2\alpha e^{-i 2 \theta_{\vb{k}}}  \\ 0 \\ -1  \\ 2 \alpha e^{i 2  \theta_{\vb{k}}} \\ 0 \\ -1 \end{pmatrix}, \nonumber \\ 
\end{equation}
for the flat bands, and 
\begin{equation}
    \psi_{\pm} ^{0} (\vb{k}) = \frac{1}{2} \begin{pmatrix} \pm e^{- i 2  \theta_{\vb{k}}} \\ - e^{- i \theta_{\vb{k}}} \\ 0 \\ \mp e^{i2 \theta_{\vb{k}}} \\  e^{i\theta_{\vb{k}}} \\ 0 \end{pmatrix},  \;\;
    \psi_{\pm} ^{1} (\vb{k}) = \frac{1}{2 \Delta_\alpha} \begin{pmatrix} e^{- i 2  \theta_{\vb{k}}} \\ \pm \Delta_\alpha e^{- i \theta_{\vb{k}}} \\ 2  \alpha \\ e^{i 2  \theta_{\vb{k}}} \\ \pm \Delta_\alpha e^{i \theta_{\vb{k}}} \\ 2\alpha  \end{pmatrix}, \nonumber
\end{equation}
\begin{figure}
    \centering
    \includegraphics[scale=1]{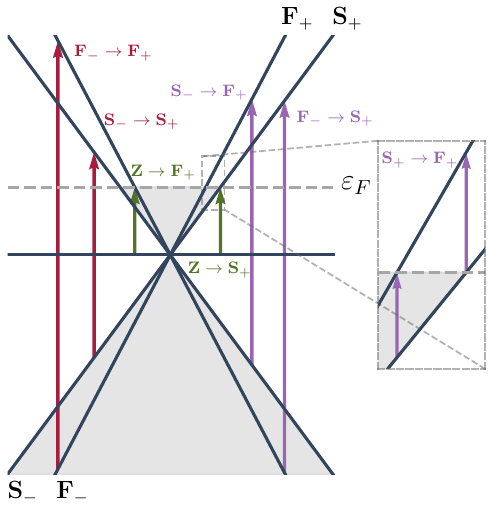}
    \caption{Schematic representation of optical interband transitions (arrows) in the double cone
    structure (blue) of the Kek-$\alpha$ model. They are categorized as intervalley (purple), intravalley (red), and flat-valley (green) transitions. Shaded regions represent filled electron states up to the Fermi energy $\varepsilon_F>0$ (gray dashed line). The inset shows the available conduction-to-conduction intervalley band transitions opened below $\varepsilon_F$ by the folding of the Brillouin zone.}
    \label{fig:fig4}
\end{figure}

\begin{figure*}
    \centering
     \includegraphics[scale=0.92]{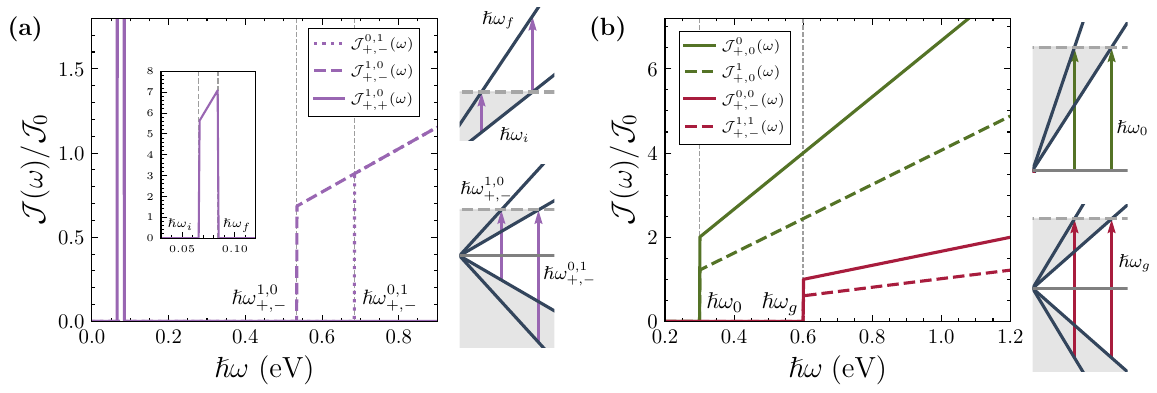}
    \caption{Joint density of states for (a) intervalley, (b) intravalley and flat-valley transitions. The inset in (a) shows the contribution of intervalley transitions between bands with the same index $\eta$ (see inset in Fig.~\ref{fig:fig4}), which occur in an energy window bounded by critical energies labeled as $\hbar\omega_i$ and $\hbar\omega_f$ (see (\ref{window-onset})).  
    The results are normalized to the JDOS between the two bands of a Dirac point at $\hbar \omega = 2|\varepsilon_F|$,
    $\J_0 = (g_s/8 \pi)(2 \vert \varepsilon_F \vert /\hslash^2 v_F ^2)$.
    We take $\varepsilon_F = 0.3$ eV and $\alpha = 0.4$.}
    \label{fig:fig5}
\end{figure*}

\noindent for the slow and fast cone, respectively. Two distinctive aspects of the present model are reflected in these eigenstates.
First, the states $\psi_0^0$ and $\psi_0^1$ are degenerate. This is due to the Brillouin zone folding induced by the Kekulé periodicity, which merges the valley states associated with the flat band. Second, only the fast states $\psi_{\pm}^{1}$ depend on the coupling parameter $\alpha$, while
the slow states $\psi_{\pm}^{0}$ remain identical to those of pristine graphene. 
This introduces an asymmetry between the fast and slow states of the nested cones
which is absent in the pure Kek-Y model, where both cones depend on the Kekulé coupling.   

\section{Optical transitions} \label{sec:secIV}

\subsection{Joint density of states} \label{subsec:secIIIA}
As a previous step to the calculation of the optical conductivity, we first explore the spectrum
of interband transitions through the joint density of states (JDOS), which for transitions $(\xi',\eta')\to (\xi,\eta)$, from the $\xi',\eta'$ band to the $\xi,\eta$ at energy $\hbar\omega$ reads as
\begin{equation} \label{eq:JDOS}
    \J_{\eta, \eta'} ^{\xi, \xi'} (\omega) = g_s \int' \frac{d^2 k}{ (2 \pi)^2} \; \delta  (\varepsilon^{\xi}_\eta (\vb{k}) - \varepsilon_{\eta'}^{\xi'} (\vb{k}) -\hslash \omega )\,,
\end{equation}
where $g_s=2$ is the spin degeneracy. 
The prime indicates an integration domain restricted to that region of $\vb{k}$-space for which $\varepsilon_{\eta'} ^{\xi'} (\vb{k}) < \varepsilon_F < \varepsilon_\eta ^\xi (\vb{k})$ (Pauli blocking), where $\varepsilon_F$ is the Fermi energy. Given that $\varepsilon^{\xi}_{\eta}({\bf k})\propto k$
(see Eq.\,(\ref{eq:energy})),
this inequality defines the radii of the wave vectors available for
the allowed transition $(\xi',\eta')\to (\xi,\eta)$ at fixed photon energy,
$\eta'\hbar v_F(\Delta_\alpha)^{\xi'}k<\varepsilon_F<\eta\hbar v_F(\Delta_\alpha)^{\xi}k$.
According to the delta function, with $\omega>0$, an additional restriction is imposed by energy conservation $\varepsilon_\eta ^\xi (\vb{k}) - \varepsilon_{\eta'} ^{\xi'} (\vb{k}) = \hslash \omega$,
which defines a circle with radius $k=\omega/(\eta v_F(\Delta_\alpha)^{\xi}-\eta' v_F(\Delta_\alpha)^{\xi'})$. The
combination of these conditions allows to find the critical energies for the interband transitions.
It can be anticipated that the JDOS will display the usual linear-in-$\omega$ dependence of
graphene-like systems.

Three sets of vertical transitions are distinguished in the present model:

(1) {\it Intravalley} transitions $(\xi,\eta'=-)\to (\xi,\eta=+)$, which we denote as
${\bf F}_-\to {\bf F}_+$ ($\xi=1$) and ${\bf S}_-\to {\bf S}_+$ ($\xi=0$). They are depicted
with red arrows in Fig.\,\ref{fig:fig4} for $\varepsilon_F>0$.

The JDOS of these transitions is
\begin{equation} \label{Intra-valley}
{\cal J}^{\xi,\xi}_{+,-}(\omega)=\frac{g_s}{8\pi}\frac{\hbar\omega}{(\hbar v_F)^2}
\frac{1}{(\Delta_{\alpha})^{2\xi}}  \ , \ \ \ \ \hbar\omega>2|\varepsilon_F|\,.
\end{equation}
Note that for the transition between slow cones ($\xi=0$), ${\bf S}_-\to {\bf S}_+$
the result is the same as for a single valley of pristine graphene. Also, when $\alpha\to 0$,
${\cal J}^{1,1}_{+,-}(\omega)={\cal J}^{0,0}_{+,-}(\omega)$ and the result for graphene
is recovered.

(2) {\it Intervalley} transitions $(\xi,\eta')\to (\Bar{\xi},\eta=+)$ [with $\Bar{\xi}=1-\xi$] for $\varepsilon_F>0$ 
(purple arrows in Fig.~\ref{fig:fig4}), 
which we refer to as \{${\bf F}_-\to {\bf S}_+,\, {\bf S}_-\to {\bf F}_+,\, {\bf S}_+\to {\bf F}_+$\},
or transitions $(\xi',\eta'=-)\to (\bar{\xi}',\eta)$ for $\varepsilon_F<0$, collected in the set
\{${\bf S}_-\to {\bf F}_+,\, {\bf F}_-\to {\bf S}_+,\, {\bf F}_-\to {\bf S}_-$\},
both sets are ordered in a decreasing energy onset sequence.
The downward transitions ${\bf F}_+\to {\bf S}_+\,$ ($\varepsilon_F>0$) and 
${\bf S}_-\to {\bf F}_-\,$ ($\varepsilon_F<0$), are excluded from the corresponding set. 
The transition ${\bf S}_+\to {\bf F}_+$ is shown in the inset of Fig.\,\ref{fig:fig4}.

\begin{figure*}
    \centering
    \includegraphics[scale=0.82]{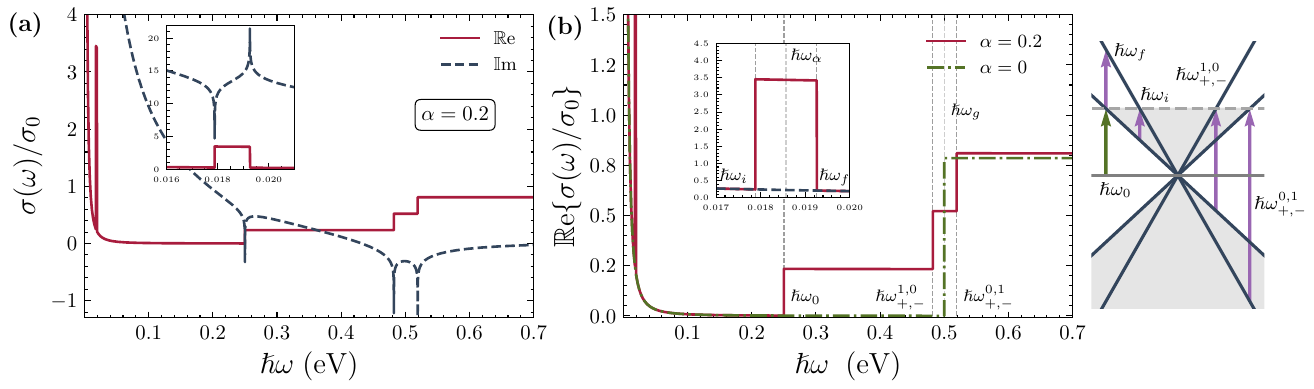}
    \caption{(a) Real and imaginary parts of the optical conductivity $\sigma (\omega)$ for 
    the Kek-$\alpha$ model. The inset shows the absorption window due to  $\re{\sigma_\alpha^{\textup{inter}} (\omega)}$ (Eq.~(\ref{sigma-alfa}))
    and the corresponding singularities in $\im{\sigma_{\alpha}^{\textup{inter}} (\omega)}$. (b) The real part of $\sigma (\omega)$ for $\alpha = 0$ (graphene) and $\alpha = 0.2$ (Kek-$\alpha$). The modulated hub atom increases and splits the interband conductivity into three steps, 
    and introduces and additional box-shaped energy window for optical absorption
    well below the Fermi energy, due to conduction-to-conduction intervalley transitions.
    The diagram on the right shows the allowed transitions contributing to the conductivity. We take $\varepsilon_F = 0.25\,$eV. }
    \label{fig:fig6}
\end{figure*}

The JDOS for the transitions with $\eta'\neq\eta$ reads as
\begin{equation} \label{interv1}
{\cal J}^{\xi,\bar{\xi}}_{+,-}(\omega)=\frac{g_s}{2\pi}\frac{\hbar\omega}{(\hbar v_F)^2}
\frac{1}{(\Delta_{\alpha}+1)^2}  \ , \ \ \ \ \hbar\omega>\left(\frac{\Delta_{\alpha}+1}{(\Delta_{\alpha})^{\xi}}\right)|\varepsilon_F|\,,
\end{equation}
while for transitions with $\eta'=\eta$,
\begin{equation}
{\cal J}^{1,0}_{+,+}(\omega)=\frac{g_s}{2\pi}\frac{\hbar\omega}{(\hbar v_F)^2}
\frac{1}{(\Delta_{\alpha}-1)^2} \,,
\end{equation}
with 
\begin{equation} \label{window-onset}
\left(\frac{\Delta_{\alpha}-1}{\Delta_{\alpha}}\right)|\varepsilon_F| <\hbar\omega<
(\Delta_{\alpha}-1)|\varepsilon_F|\,.
\end{equation}
The appearance of this window below $|\varepsilon_F|$ is a clear signature of the present hybrid model; as $\alpha \rightarrow 0$, it leads to the formation of a singularity, an effect known as band nesting \cite{CarvalhoBandNesting2013, mennel2020band}.
Later, we will see how this impacts the conductivity and 
serves as a distinctive feature of Kekulé periodicity.

(3) {\it Flat-valley} transitions ($\xi',\eta'=0) \to (\xi,\eta=+)$ ($\varepsilon_F>0$) or
$(\xi',\eta'=-)\to (\xi,\eta=0)$ ($\varepsilon_F<0$), denoted as \{${\bf Z}^{\xi'}\to {\bf F}_+$, 
${\bf Z}^{\xi'}\to {\bf S}_+$\} (green arrows in Fig.\,\ref{fig:fig4})
or \{${\bf F}_-\to {\bf Z}^\xi$, ${\bf S}_-\to {\bf Z}^\xi$\}, respectively.
For the former set, Eq.\,(\ref{eq:JDOS}) with $\varepsilon^{\xi'}_{\eta'}=0$, gives
\begin{equation} \label{flat-valley}
{\cal J}^{\xi}_{+,0}(\omega)=\frac{g_s}{2\pi}\frac{\hbar\omega}{(\hbar v_F)^2}
\frac{1}{(\Delta_{\alpha})^{2\xi}}  \ , \ \ \ \ \hbar\omega>|\varepsilon_F|\,.   
\end{equation}
For the latter set, taking $\varepsilon^{\xi}_{\eta}=0$ in (\ref{eq:JDOS}), ${\cal J}^{\xi'}_{0,-}(\omega)={\cal J}^{\xi'}_{+,0}(\omega)$. 

Figure \ref{fig:fig5} displays the JDOS for the complete set of interband transitions of the
present model. The corresponding onsets in Eqs. (\ref{Intra-valley})-(\ref{flat-valley})
have been labeled according to the definition
\begin{equation}
\hbar\omega^{\xi,\xi'}_{\eta,\eta'}=(\eta-\eta'(\Delta_{\alpha})^{\xi'-\xi})
|\varepsilon_F| \,.  
\end{equation}
Thus, $\hslash \omega_g \equiv \hbar\omega^{1,1}_{+,-}=\hbar\omega^{0,0}_{+,-}=2|\varepsilon_F|$ corresponds to
the onset for the intravalley transitions, while $\hbar\omega^{0,1}_{+,-}=(\Delta_{\alpha}+1)|\varepsilon_F|$,
$\hbar\omega^{1,0}_{+,-}=(\Delta_{\alpha}+1)|\varepsilon_F|/\Delta_{\alpha}$ (see (\ref{interv1})), 
and
$\hbar\omega^{1,0}_{+,+}=(\Delta_{\alpha}-1)|\varepsilon_F|/\Delta_{\alpha}\equiv \hbar\omega_i$, 
$\Delta_{\alpha}\hbar\omega^{1,0}_{+,+}=(\Delta_{\alpha}-1)|\varepsilon_F|\equiv\hbar\omega_f$ 
(see (\ref{window-onset})), to the intervalley transitions. The corresponding energy onset for
the flat-valley transitions are labeled as $\hbar\omega_0\equiv\hbar\omega^{\xi,\xi'}_{+,0}=
\hbar\omega^{\xi',\xi'}_{0,-}=|\varepsilon_F|$.

It can be seen how for transitions sharing the onset, the $\alpha$-dependent slope provides a way to 
identify its nature.
It is worthwhile to note also that for a decreasing magnitude of the parameter $\alpha$,
the number of transitions between cones with the same band index, ${\bf S}_+\to {\bf F}_+$
or ${\bf F}_-\to {\bf S}_-$, notably increase because  
${\cal J}^{+,-}_{+,+}(\omega)\propto(\Delta_{\alpha}-1)^{-2}$, although the frequency region 
(\ref{window-onset}) narrows. 

\subsection{Optical conductivity} \label{subsec:secIIIB}

The optical conductivity tensor of the system reduces to a scalar response function,
with real and imaginary parts
\begin{align}
     \re{\sigma (\omega)} & = D \delta (\omega) + \re{\sigma ^{\textup{inter}} (\omega)} \label{eq:realpart}\,,  \\
     \im{\sigma (\omega)} & = \im{\sigma ^{\textup{intra}} (\omega)} + \im{\sigma ^{\textup{inter}} (\omega)}\,,  
\end{align}
where the intraband and interband contributions are obtained from the current-current Kubo
formula as
\begin{equation}
    \sigma ^{\textup{intra}} (\omega) = i g_s  \frac{\sigma_0}{4 \pi \hslash \omega} \sum_{\xi,\eta} \int d^2 k \; \delta \left(\varepsilon_{\eta} ^{\xi} -\varepsilon_F \right) \left( \frac{\partial \varepsilon_\eta ^{\xi}}{\partial k_x} \right)^2,
\end{equation}
\begin{align} \label{eq:realcon}
      \re{\sigma ^{\textup{inter}} (\omega)} & = g_s\sigma_0\frac{ \hbar\omega}{4} \sum_{\substack{\xi,\xi',\eta, \eta'}} 
      \int'\! d^2 k \; V_{\eta, \eta'} ^{\xi, \xi'} (\vb{k})  \\
      &  \hspace*{2cm} \times 
      \; \delta \left( \varepsilon^{\xi}_{\eta}({\bf k})-\varepsilon^{\xi'}_{\eta'}({\bf k}) - \hslash \omega \right)  , \nonumber
\end{align}
\begin{align} \label{eq:imagcon} 
       \im{\sigma ^{\textup{inter}} (\omega)}  & =  g_s \sigma_0 \frac{\hslash \omega}{2 \pi}  \sum_{\substack{\xi,\xi',\eta, \eta'}} 
       \mathcal{P}\!\! \int' \!d^2 k \; V_{\eta, \eta'} ^{\xi, \xi'} (\vb{k}) \\ 
       & \hspace*{1.5cm}\; \times \frac{\varepsilon^{\xi}_{\eta}({\bf k})-\varepsilon^{\xi'}_{\eta'}({\bf k})} { ( \hslash \omega)^2 - \left(\varepsilon^{\xi}_{\eta}({\bf k})-\varepsilon^{\xi'}_{\eta'}({\bf k})\right)^2   } , \nonumber
\end{align}
assuming zero temperature. Here, $\sigma_0 = 2 e^2 /h$ and $\mathcal{P}$ denotes Principal Value integral. 
The function $V_{\eta, \eta'} ^{\xi, \xi'} (\vb{k})$ arises from the product of matrix elements of the velocity operator or in terms of the interband Berry connection as $V_{\eta, \eta'} ^{\xi, \xi'} (\vb{k}) = \mathcal{A}_{\eta, \eta'} ^{\xi, \xi'}   (\vb{k})  \mathcal{A}_{\eta', \eta} ^{\xi', \xi}   (\vb{k}) $ where $\mathcal{A}_{\eta, \eta'} ^{\xi, \xi'} (\vb{k}) = i \braket{\psi_\eta ^{\xi} (\vb{k}) \vert \partial_{k_i} \psi_{\eta'} ^{\xi'} (\vb{k}) }$ is the $k_i$-component, with $i=\{x, y\}$, of the interband Berry connection. We choose $i=x$ for concreteness given the isotropy of the model.
The prime in the integrals demands the same restriction as in Eq.\,\eqref{eq:JDOS}. We have included in Eq.~(\ref{eq:realpart})  the Drude weight $D = \pi \lim_{\omega \rightarrow 0} [ \omega \im{\sigma^{\textup{intra}} (\omega)}  ]$. 

The elements $V_{\eta, \eta'} ^{\xi, \xi'} (\vb{k})$ are given by:
\begin{equation}
    V_{\eta, \eta'} ^{\xi, \xi'} (\vb{k}) = \frac{ \mathcal{V}^{\xi,\xi'} _{\eta,\eta'}}{ [\varepsilon^{\xi}_{\eta}({\bf k})-\varepsilon^{\xi'}_{\eta'}({\bf k})]^2 } ,
\end{equation}
where we have defined the dimensionless coefficients $\mathcal{V}^{\xi,\xi'} _{\eta,\eta'}= \bra{\psi_\eta^\xi}  \partial_{k_x} \Ham  \ket{\psi_{\eta'}^{\xi '}}\bra{\psi_{\eta'}^{\xi '}}  \partial_{k_x} \Ham  \ket{\psi_\eta^\xi}$. For allowed interband transitions, we have  
$[\eta \eta' (4 \alpha^2 - 1) + \Delta_\alpha]^2 / 4 \Delta_{\alpha}^{2} $ for intervalley transitions, $4 \alpha^2 / (2\alpha^2 +1) $ for the transition $\textbf{Z}^1 \to \textbf{S}_+$ (or $\textbf{S}_- \to \textbf{Z}^1$) and zero otherwise. 

Intravalley transitions ($\textbf{S}_- \rightarrow \textbf{S}_+$ and $\textbf{F}_- \rightarrow \textbf{F}_+$) in the Kek-Y graphene model were previously shown to be forbidden using Fermi’s golden rule~\cite{HerreraElectronicOpticalKekY}. This can be attributed to the fact that the $\vb{S}$ cone is entirely chiral, while the $\vb{F}$ cone is entirely antichiral. A more formal verification was provided using symmetry arguments, which impose a selection rule in the context of the Zitterbewegung effect~\cite{ValleydrivenSantacruz2022}. In addition, we find that the transitions $\vb{Z}^\xi \to \vb{F}_{+}$, between a flat band and the fast cone are absent, reducing the number of transitions with the flat band, compared to the $\Keka$ model.

The total conductivity has intraband and interband contributions, \(\sigma (\omega)=\sigma^{\textup{intra}}(\omega)+\sigma^{\textup{inter}}(\omega)\). The intraband conductivity, as in pristine graphene~\cite{MakMeasurementOpticalGraphene2008, LAFalkovsky2008, HorngDrudeCondGraphene2011,Mojarro2020}, is given by
\begin{equation}
    \sigma ^{\textup{intra}} (\omega) = g_s \sigma_0  \frac{\pi}{2} \left[ \delta (\hslash \omega) + i \frac{1}{ \pi \hslash \omega}  \right] |\varepsilon_F|,
\end{equation}
where each cone contributed independently and the  flat band did not, since it has uniformly zero group velocity.

We divide the interband contribution as 
\begin{equation}
    \sigma^{\textup{inter}} (\omega) = \sigma^{\textup{inter}}_{>} (\omega) \Theta(\hbar\omega-\vert \varepsilon_F \vert)+ \sigma^{\textup{inter}}_{<}  (\omega)\Theta(\vert \varepsilon_F \vert -\hbar\omega) 
\end{equation}
where $\sigma^{\textup{inter}}_{>}  (\omega)$ considers contributions above the Fermi energy, 
that is the sum of all permitted interband transitions, excluding \(\textbf{S}_+ \rightarrow \textbf{F}_+\)
(or $\textbf{F}_- \rightarrow \textbf{S}_-$ when $\varepsilon_F<0$), therefore
\begin{equation}
    \sigma_{>} ^{\textup{inter}} (\omega) = [ \sigma_{>} ^{\textup{inter}} (\omega) ]_{+,0} ^{0,1} + [ \sigma_{>} ^{\textup{inter}} (\omega) ]_{+,-} ^{1,0} +[ \sigma_{>} ^{\textup{inter}} (\omega) ]_{+,-} ^{0,1}
\end{equation}
with each transition contribution given by,
\begin{align} \label{eq:intercond}
    & \left[\sigma_> ^{\textup{inter}} (\omega) \right]_{\eta, \eta'} ^{\xi, \xi'} =  \; g_s \sigma_{0} \frac{\pi}{2\hbar^2v_F^2} \frac{\mathcal{V}_{\eta, \eta'} ^{\xi, \xi'} }{(\eta (\Delta_\alpha )^\xi - \eta' (\Delta_\alpha) ^{\xi'})^2}     \nonumber \\
    & \times \left\lbrace \Theta (\hslash \omega - \hslash \omega _{\eta, \eta'} ^{\xi, \xi'} ) - \frac{i}{\pi} \ln \left[ \frac{\vert \hslash \omega + \hslash \omega_{\eta, \eta'} ^{\xi, \xi'} \vert}{ \vert \hslash \omega - \hslash \omega_{\eta, \eta'} ^{\xi, \xi'} \vert} \right] \right\rbrace .
\end{align}

\begin{figure}
    \centering
    \includegraphics[scale=1.05]{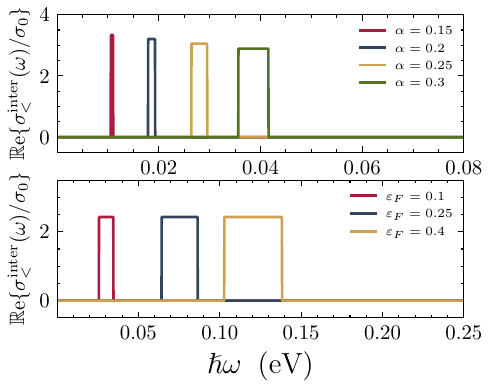}
    \caption{Top: $\re{\sigma_<^{\textup{inter}}(\omega)}$ for several values of $\alpha$ at
    fixed $\varepsilon_F=0.25\,$eV. The absorption window displays a red shift and narrows as parameter $\alpha$ decreases, with its magnitude increasing (see Eq.~(\ref{sigma-alfa})).
    Bottom: $\re{\sigma_\alpha^{\textup{inter}}(\omega)}$ for varying $\varepsilon_F$ 
    with fixed value of $\alpha=0.45$. A red shift and a  narrowing of the spectrum is observed
    as $\varepsilon_F$ diminish, but now its magnitude remains constant.}
    \label{fig:fig7}
\end{figure}

On the other hand,  $\sigma^{\textup{inter}}_{<}  (\omega)$ considers contributions below the Fermi energy, that is, the intervalley transitions \(\textbf{S}_+ \rightarrow \textbf{F}_+\) and $\textbf{F}_- \rightarrow \textbf{S}_-$, then  
\begin{align} \label{sigma-alfa}
    \sigma_{<} ^{\textup{inter}} (\omega) = & \;g_s \frac{\sigma_0\pi}{2\hbar^2v_F^2}  \frac{\mathcal{V}_{+, +} ^{1, 0}  }{( \Delta_{\alpha}-1)^2} \left\lbrace \Theta \left(\hslash \omega  - \hslash \omega_i \right) \Theta \left(\hslash \omega_f - \hslash \omega \right) \right. \nonumber \\
    & \left. \hspace*{1cm}  + \frac{i}{\pi} \ln \left| \frac{\hslash \omega -  \hslash \omega_f }{\hslash \omega +  \hslash \omega_f }\cdot \frac{\hslash \omega + \hslash \omega_i }{\hslash \omega - \hslash \omega_i }     \right|\right\rbrace .
\end{align}

The real and imaginary parts of the total optical conductivity $\sigma(\omega)$, for $\varepsilon_F=0.25$~eV, are shown in Fig.~\hyperref[fig:fig6]{6(a)}, in Fig.~\hyperref[fig:fig6]{6(b)} we compare the real part of the conductivity for $\alpha=0$ (graphene) and $\alpha=0.4$. 
Compared with the usual optical response of graphene, these plots show three distinctive features: an intervalley  window for transitions below the Fermi energy (see the inset in Fig.~\hyperref[fig:fig6]{6(a)}), a flat band to slow cone step at $\hbar\omega=\varepsilon_F$ and the splitting of the step at $\hbar\omega=2\varepsilon_F$ in two steps, one at 
$\hbar\omega^{0,1}_{+,-}=(\Delta_{\alpha}+1)|\varepsilon_F|$, and the other at
$\hbar\omega^{1,0}_{+,-}=(\Delta_{\alpha}+1)|\varepsilon_F|/\Delta_{\alpha}$. Notably, the addition of the hub atom results in increased maximum absorption compared to that of other graphene-like systems, with this distinct three-step structure in the optical conductivity~\cite{HerreraElectronicOpticalKekY, HerreraDynamicPlasmonsKekule2020}. This effect, along with the modified energy dispersion, resembles the behavior of a 5/2-pseudospin Dirac semimetal~\cite{DoraLatticeGeneralization2011}, suggesting that lattice modulation may alter the system's effective pseudospin.

When $\alpha \rightarrow 0$,  the $\sigma^{\textup{inter}} _< (\omega)$ contribution and the flat-band term in Eq.~(\ref{eq:intercond}) vanish, reducing the interband conductivity to the well-known expression for pristine graphene's interband conductivity~\cite{LAFalkovsky2008}, $\sigma^{\textup{inter}} (\omega)=\sigma_{\text{gr}}^{\textup{inter}} (\omega)$, where
\begin{align}
    \sigma_{\text{gr}}^{\textup{inter}} (\omega) = g_s \frac{\sigma_0\pi}{16} \left\lbrace \Theta (\hslash \omega - 2  |\varepsilon_F| )  - \frac{i}{\pi} \ln \left| \frac{ \hslash \omega + 2 |\varepsilon_F|}{  \hslash \omega - 2 |\varepsilon_F| } \right| \right\rbrace .
\end{align}

We further analyze the interband contributions below the Fermi energy  $\sigma_< ^{\textup{inter}} (\omega)$ in Fig.\ref{fig:fig7}.
The frequency range of this conductivity is determined by $\hslash \omega_{fi}  = \hslash( \omega_f - \omega_i )  = (\Delta_\alpha  -  1)^2    \vert \varepsilon_F \vert / \Delta_\alpha$, and 
it is centered at $\hslash \omega_< = \hbar(\omega_i+\omega_f)/2 =(\Delta_\alpha ^2 - 1)\vert \varepsilon_F \vert/(2 \Delta_\alpha)$, as shown in the inset of Fig.~\hyperref[fig:fig6]{6(b)}. Notice that  $\omega_<= (\omega_{+,-} ^{0,1} - \omega_{+,-} ^{1,0} )/2 $, i.e. $\omega_<$ is related with the frequency difference of the transitions \(\textbf{S}_- \rightarrow \textbf{F}_+\) and $\textbf{F}_- \rightarrow \textbf{S}_+$.
The energy  $\hslash \omega_<$ is often referred to as the beat frequency~\cite{HerreraDynamicPlasmonsKekule2020} and is the result of interference between the two closely spaced critical frequencies.
For $\alpha \to 0$, the  contribution  $\sigma_< ^{\textup{inter}} (\omega)$ vanishes, as the band $\varepsilon_+ ^+ (\vb{k})$ and $\varepsilon_+ ^- (\vb{k})$ nearly overlap, resulting in $\nabla_{\vb{k}} [\varepsilon_+ ^+ (\vb{k}) - \varepsilon_+ ^- (\vb{k})] \approx 0$ and resulting in an absorption peak in conductivity. This effect, known as band nesting, has also been observed in transition metal dichalcogenides~\cite{CarvalhoBandNesting2013, mennel2020band} and space-modulated 2D materials, such as twisted bilayer graphene~\cite{MoirephononsKoshino2019, OchoaFlatBandsMoire2020, cao2018unconventional}.

\section{Conclusions} \label{sec:secV}

In this work, we study the effects of atoms appearing with Kekulé periodicity $(\sqrt{3} \times \sqrt{3})$ on a honeycomb lattice, coupling to one of the sublattices. This creates a hybrid model combining features of the $\Keka$ model and Kekulé-distorted graphene, which we call Kekulé-modulated $\Keka$ model or Kek-$\alpha$.

We calculated the band structure and corresponding eigenfunctions, featuring a double-cone structure with a degenerate flat band, closely resembling the Kek-Y model. Furthermore, we studied the optical transitions through the joint density of states (JDOS) and optical conductivity. Notably, compared to the $\Keka$ model, new terms in the conductivity emerge due to the opening of intervalley channels, which are absent in both the $\Keka$ model and Kekulé-distorted graphene, and manifest as critical frequencies in the optical spectra. In contrast to the Kek-Y graphene model, the optical response of our model demonstrates significant tunability through the Kekulé parameter $\alpha$. In the low-energy approximation, our analytical results align with those of pristine graphene under appropriate limits. Finally, we describe an absorption phenomenon characterized by a resonance frequency linked to intervalley transport, which appears at a beat frequency determined by the characteristic frequencies of each valley. This behavior also occurs in the Kek-Y model, indicating that this resonance frequency may serve as a reliable signature for identifying Kekulé periodicity in similar systems.

\section*{Acknowledgments}

L.E.S. thanks CONAHCYT for a MSc scholarship. M.A.M. acknowledges U.S. Department of Energy, Office of Basic Energy Sciences, Materials Science and Engineering
Division.

\bibliography{references.bib}

\begin{thebibliography}{80}%
\makeatletter
\providecommand \@ifxundefined [1]{%
 \@ifx{#1\undefined}
}%
\providecommand \@ifnum [1]{%
 \ifnum #1\expandafter \@firstoftwo
 \else \expandafter \@secondoftwo
 \fi
}%
\providecommand \@ifx [1]{%
 \ifx #1\expandafter \@firstoftwo
 \else \expandafter \@secondoftwo
 \fi
}%
\providecommand \natexlab [1]{#1}%
\providecommand \enquote  [1]{``#1''}%
\providecommand \bibnamefont  [1]{#1}%
\providecommand \bibfnamefont [1]{#1}%
\providecommand \citenamefont [1]{#1}%
\providecommand \href@noop [0]{\@secondoftwo}%
\providecommand \href [0]{\begingroup \@sanitize@url \@href}%
\providecommand \@href[1]{\@@startlink{#1}\@@href}%
\providecommand \@@href[1]{\endgroup#1\@@endlink}%
\providecommand \@sanitize@url [0]{\catcode `\\12\catcode `\$12\catcode `\&12\catcode `\#12\catcode `\^12\catcode `\_12\catcode `\%12\relax}%
\providecommand \@@startlink[1]{}%
\providecommand \@@endlink[0]{}%
\providecommand \url  [0]{\begingroup\@sanitize@url \@url }%
\providecommand \@url [1]{\endgroup\@href {#1}{\urlprefix }}%
\providecommand \urlprefix  [0]{URL }%
\providecommand \Eprint [0]{\href }%
\providecommand \doibase [0]{https://doi.org/}%
\providecommand \selectlanguage [0]{\@gobble}%
\providecommand \bibinfo  [0]{\@secondoftwo}%
\providecommand \bibfield  [0]{\@secondoftwo}%
\providecommand \translation [1]{[#1]}%
\providecommand \BibitemOpen [0]{}%
\providecommand \bibitemStop [0]{}%
\providecommand \bibitemNoStop [0]{.\EOS\space}%
\providecommand \EOS [0]{\spacefactor3000\relax}%
\providecommand \BibitemShut  [1]{\csname bibitem#1\endcsname}%
\let\auto@bib@innerbib\@empty
\bibitem [{\citenamefont {Kopnin}\ \emph {et~al.}(2011)\citenamefont {Kopnin}, \citenamefont {Heikkil\"a},\ and\ \citenamefont {Volovik}}]{KopninSCTopologicalFlatBand2011}%
  \BibitemOpen
  \bibfield  {author} {\bibinfo {author} {\bibfnamefont {N.~B.}\ \bibnamefont {Kopnin}}, \bibinfo {author} {\bibfnamefont {T.~T.}\ \bibnamefont {Heikkil\"a}},\ and\ \bibinfo {author} {\bibfnamefont {G.~E.}\ \bibnamefont {Volovik}},\ }\bibfield  {title} {\bibinfo {title} {High-temperature surface superconductivity in topological flat-band systems},\ }\href {https://doi.org/10.1103/PhysRevB.83.220503} {\bibfield  {journal} {\bibinfo  {journal} {Phys. Rev. B}\ }\textbf {\bibinfo {volume} {83}},\ \bibinfo {pages} {220503} (\bibinfo {year} {2011})}\BibitemShut {NoStop}%
\bibitem [{\citenamefont {Cao}\ \emph {et~al.}(2018)\citenamefont {Cao}, \citenamefont {Fatemi}, \citenamefont {Fang}, \citenamefont {Watanabe}, \citenamefont {Taniguchi}, \citenamefont {Kaxiras},\ and\ \citenamefont {Jarillo-Herrero}}]{cao2018unconventional}%
  \BibitemOpen
  \bibfield  {author} {\bibinfo {author} {\bibfnamefont {Y.}~\bibnamefont {Cao}}, \bibinfo {author} {\bibfnamefont {V.}~\bibnamefont {Fatemi}}, \bibinfo {author} {\bibfnamefont {S.}~\bibnamefont {Fang}}, \bibinfo {author} {\bibfnamefont {K.}~\bibnamefont {Watanabe}}, \bibinfo {author} {\bibfnamefont {T.}~\bibnamefont {Taniguchi}}, \bibinfo {author} {\bibfnamefont {E.}~\bibnamefont {Kaxiras}},\ and\ \bibinfo {author} {\bibfnamefont {P.}~\bibnamefont {Jarillo-Herrero}},\ }\bibfield  {title} {\bibinfo {title} {Unconventional superconductivity in magic-angle graphene superlattices},\ }\href {https://doi.org/10.1038/nature26160} {\bibfield  {journal} {\bibinfo  {journal} {Nature}\ }\textbf {\bibinfo {volume} {556}},\ \bibinfo {pages} {43} (\bibinfo {year} {2018})}\BibitemShut {NoStop}%
\bibitem [{\citenamefont {Ehlen}\ \emph {et~al.}(2020)\citenamefont {Ehlen}, \citenamefont {Hell}, \citenamefont {Marini}, \citenamefont {Hasdeo}, \citenamefont {Saito}, \citenamefont {Falke}, \citenamefont {Goerbig}, \citenamefont {Di~Santo}, \citenamefont {Petaccia}, \citenamefont {Profeta},\ and\ \citenamefont {Grüneis}}]{EhlenOriginFlatBand2020}%
  \BibitemOpen
  \bibfield  {author} {\bibinfo {author} {\bibfnamefont {N.}~\bibnamefont {Ehlen}}, \bibinfo {author} {\bibfnamefont {M.}~\bibnamefont {Hell}}, \bibinfo {author} {\bibfnamefont {G.}~\bibnamefont {Marini}}, \bibinfo {author} {\bibfnamefont {E.~H.}\ \bibnamefont {Hasdeo}}, \bibinfo {author} {\bibfnamefont {R.}~\bibnamefont {Saito}}, \bibinfo {author} {\bibfnamefont {Y.}~\bibnamefont {Falke}}, \bibinfo {author} {\bibfnamefont {M.~O.}\ \bibnamefont {Goerbig}}, \bibinfo {author} {\bibfnamefont {G.}~\bibnamefont {Di~Santo}}, \bibinfo {author} {\bibfnamefont {L.}~\bibnamefont {Petaccia}}, \bibinfo {author} {\bibfnamefont {G.}~\bibnamefont {Profeta}},\ and\ \bibinfo {author} {\bibfnamefont {A.}~\bibnamefont {Grüneis}},\ }\bibfield  {title} {\bibinfo {title} {Origin of the flat band in heavily cs-doped graphene},\ }\href {https://doi.org/10.1021/acsnano.9b08622} {\bibfield  {journal} {\bibinfo  {journal} {ACS Nano}\ }\textbf {\bibinfo {volume} {14}},\ \bibinfo {pages} {1055} (\bibinfo {year} {2020})}\BibitemShut
  {NoStop}%
\bibitem [{\citenamefont {Leykam}\ \emph {et~al.}(2018)\citenamefont {Leykam}, \citenamefont {Andreanov},\ and\ \citenamefont {Flach}}]{Leykam2018ArtificialExperiments}%
  \BibitemOpen
  \bibfield  {author} {\bibinfo {author} {\bibfnamefont {D.}~\bibnamefont {Leykam}}, \bibinfo {author} {\bibfnamefont {A.}~\bibnamefont {Andreanov}},\ and\ \bibinfo {author} {\bibfnamefont {S.}~\bibnamefont {Flach}},\ }\bibfield  {title} {\bibinfo {title} {{Artificial flat band systems: from lattice models to experiments}},\ }\href {https://doi.org/10.1080/23746149.2018.1473052} {\bibfield  {journal} {\bibinfo  {journal} {Advances in Physics: X}\ }\textbf {\bibinfo {volume} {3}},\ \bibinfo {pages} {1473052} (\bibinfo {year} {2018})}\BibitemShut {NoStop}%
\bibitem [{\citenamefont {Drost}\ \emph {et~al.}(2017)\citenamefont {Drost}, \citenamefont {Ojanen}, \citenamefont {Harju},\ and\ \citenamefont {Liljeroth}}]{drost2017topologicalstates}%
  \BibitemOpen
  \bibfield  {author} {\bibinfo {author} {\bibfnamefont {R.}~\bibnamefont {Drost}}, \bibinfo {author} {\bibfnamefont {T.}~\bibnamefont {Ojanen}}, \bibinfo {author} {\bibfnamefont {A.}~\bibnamefont {Harju}},\ and\ \bibinfo {author} {\bibfnamefont {P.}~\bibnamefont {Liljeroth}},\ }\bibfield  {title} {\bibinfo {title} {Topological states in engineered atomic lattices},\ }\href {https://doi.org/10.1038/nphys4080} {\bibfield  {journal} {\bibinfo  {journal} {Nature Physics}\ }\textbf {\bibinfo {volume} {13}},\ \bibinfo {pages} {668} (\bibinfo {year} {2017})}\BibitemShut {NoStop}%
\bibitem [{\citenamefont {Al~Ezzi}\ \emph {et~al.}(2024)\citenamefont {Al~Ezzi}, \citenamefont {Hu}, \citenamefont {Ariando}, \citenamefont {Guinea},\ and\ \citenamefont {Adam}}]{AlEzziTopologicalFlatBand2024}%
  \BibitemOpen
  \bibfield  {author} {\bibinfo {author} {\bibfnamefont {M.~M.}\ \bibnamefont {Al~Ezzi}}, \bibinfo {author} {\bibfnamefont {J.}~\bibnamefont {Hu}}, \bibinfo {author} {\bibfnamefont {A.}~\bibnamefont {Ariando}}, \bibinfo {author} {\bibfnamefont {F.}~\bibnamefont {Guinea}},\ and\ \bibinfo {author} {\bibfnamefont {S.}~\bibnamefont {Adam}},\ }\bibfield  {title} {\bibinfo {title} {Topological flat bands in graphene super-moir\'e lattices},\ }\href {https://doi.org/10.1103/PhysRevLett.132.126401} {\bibfield  {journal} {\bibinfo  {journal} {Phys. Rev. Lett.}\ }\textbf {\bibinfo {volume} {132}},\ \bibinfo {pages} {126401} (\bibinfo {year} {2024})}\BibitemShut {NoStop}%
\bibitem [{\citenamefont {Bao}\ \emph {et~al.}(2022)\citenamefont {Bao}, \citenamefont {Zhang}, \citenamefont {Wu}, \citenamefont {Zhou}, \citenamefont {Li}, \citenamefont {Yu}, \citenamefont {Li}, \citenamefont {Duan},\ and\ \citenamefont {Zhou}}]{BaoFlatBand2022}%
  \BibitemOpen
  \bibfield  {author} {\bibinfo {author} {\bibfnamefont {C.}~\bibnamefont {Bao}}, \bibinfo {author} {\bibfnamefont {H.}~\bibnamefont {Zhang}}, \bibinfo {author} {\bibfnamefont {X.}~\bibnamefont {Wu}}, \bibinfo {author} {\bibfnamefont {S.}~\bibnamefont {Zhou}}, \bibinfo {author} {\bibfnamefont {Q.}~\bibnamefont {Li}}, \bibinfo {author} {\bibfnamefont {P.}~\bibnamefont {Yu}}, \bibinfo {author} {\bibfnamefont {J.}~\bibnamefont {Li}}, \bibinfo {author} {\bibfnamefont {W.}~\bibnamefont {Duan}},\ and\ \bibinfo {author} {\bibfnamefont {S.}~\bibnamefont {Zhou}},\ }\bibfield  {title} {\bibinfo {title} {Coexistence of extended flat band and kekul\'e order in li-intercalated graphene},\ }\href {https://doi.org/10.1103/PhysRevB.105.L161106} {\bibfield  {journal} {\bibinfo  {journal} {Phys. Rev. B}\ }\textbf {\bibinfo {volume} {105}},\ \bibinfo {pages} {L161106} (\bibinfo {year} {2022})}\BibitemShut {NoStop}%
\bibitem [{\citenamefont {Escudero}\ \emph {et~al.}(2024)\citenamefont {Escudero}, \citenamefont {Sinner}, \citenamefont {Zhan}, \citenamefont {Pantale\'on},\ and\ \citenamefont {Guinea}}]{EscuderoDesignigMoireStrain2024}%
  \BibitemOpen
  \bibfield  {author} {\bibinfo {author} {\bibfnamefont {F.}~\bibnamefont {Escudero}}, \bibinfo {author} {\bibfnamefont {A.}~\bibnamefont {Sinner}}, \bibinfo {author} {\bibfnamefont {Z.}~\bibnamefont {Zhan}}, \bibinfo {author} {\bibfnamefont {P.~A.}\ \bibnamefont {Pantale\'on}},\ and\ \bibinfo {author} {\bibfnamefont {F.}~\bibnamefont {Guinea}},\ }\bibfield  {title} {\bibinfo {title} {Designing moir\'e patterns by strain},\ }\href {https://doi.org/10.1103/PhysRevResearch.6.023203} {\bibfield  {journal} {\bibinfo  {journal} {Phys. Rev. Res.}\ }\textbf {\bibinfo {volume} {6}},\ \bibinfo {pages} {023203} (\bibinfo {year} {2024})}\BibitemShut {NoStop}%
\bibitem [{\citenamefont {de~Jes{\'u}s Espinosa-Champo}\ and\ \citenamefont {Naumis}(2024)}]{de2024flat}%
  \BibitemOpen
  \bibfield  {author} {\bibinfo {author} {\bibfnamefont {A.}~\bibnamefont {de~Jes{\'u}s Espinosa-Champo}}\ and\ \bibinfo {author} {\bibfnamefont {G.~G.}\ \bibnamefont {Naumis}},\ }\bibfield  {title} {\bibinfo {title} {Flat bands without twists: periodic holey graphene},\ }\href {https://doi.org/10.1088/1361-648X/ad39be} {\bibfield  {journal} {\bibinfo  {journal} {Journal of Physics: Condensed Matter}\ }\textbf {\bibinfo {volume} {36}},\ \bibinfo {pages} {275703} (\bibinfo {year} {2024})}\BibitemShut {NoStop}%
\bibitem [{\citenamefont {Deng}\ \emph {et~al.}(2003)\citenamefont {Deng}, \citenamefont {Simon},\ and\ \citenamefont {K{\"o}hler}}]{deng2003origin}%
  \BibitemOpen
  \bibfield  {author} {\bibinfo {author} {\bibfnamefont {S.}~\bibnamefont {Deng}}, \bibinfo {author} {\bibfnamefont {A.}~\bibnamefont {Simon}},\ and\ \bibinfo {author} {\bibfnamefont {J.}~\bibnamefont {K{\"o}hler}},\ }\bibfield  {title} {\bibinfo {title} {The origin of a flat band},\ }\href {https://doi.org/10.1016/S0022-4596(03)00239-1} {\bibfield  {journal} {\bibinfo  {journal} {Journal of Solid State Chemistry}\ }\textbf {\bibinfo {volume} {176}},\ \bibinfo {pages} {412} (\bibinfo {year} {2003})}\BibitemShut {NoStop}%
\bibitem [{\citenamefont {Roman-Taboada}\ and\ \citenamefont {Naumis}(2017)}]{TopologicalFlatBandTaboada2017}%
  \BibitemOpen
  \bibfield  {author} {\bibinfo {author} {\bibfnamefont {P.}~\bibnamefont {Roman-Taboada}}\ and\ \bibinfo {author} {\bibfnamefont {G.~G.}\ \bibnamefont {Naumis}},\ }\bibfield  {title} {\bibinfo {title} {Topological flat bands in time-periodically driven uniaxial strained graphene nanoribbons},\ }\href {https://doi.org/10.1103/PhysRevB.95.115440} {\bibfield  {journal} {\bibinfo  {journal} {Phys. Rev. B}\ }\textbf {\bibinfo {volume} {95}},\ \bibinfo {pages} {115440} (\bibinfo {year} {2017})}\BibitemShut {NoStop}%
\bibitem [{\citenamefont {Bistritzer}\ and\ \citenamefont {MacDonald}(2011)}]{MacDonaldMoireBands2011}%
  \BibitemOpen
  \bibfield  {author} {\bibinfo {author} {\bibfnamefont {R.}~\bibnamefont {Bistritzer}}\ and\ \bibinfo {author} {\bibfnamefont {A.~H.}\ \bibnamefont {MacDonald}},\ }\bibfield  {title} {\bibinfo {title} {Moir{\'e} bands in twisted double-layer graphene},\ }\href {https://doi.org/10.1073/pnas.1108174108} {\bibfield  {journal} {\bibinfo  {journal} {Proceedings of the National Academy of Sciences}\ }\textbf {\bibinfo {volume} {108}},\ \bibinfo {pages} {12233} (\bibinfo {year} {2011})}\BibitemShut {NoStop}%
\bibitem [{\citenamefont {Mogera}\ and\ \citenamefont {Kulkarni}(2020)}]{mogera2020new}%
  \BibitemOpen
  \bibfield  {author} {\bibinfo {author} {\bibfnamefont {U.}~\bibnamefont {Mogera}}\ and\ \bibinfo {author} {\bibfnamefont {G.~U.}\ \bibnamefont {Kulkarni}},\ }\bibfield  {title} {\bibinfo {title} {A new twist in graphene research: Twisted graphene},\ }\href {https://doi.org/10.1016/j.carbon.2019.09.053} {\bibfield  {journal} {\bibinfo  {journal} {Carbon}\ }\textbf {\bibinfo {volume} {156}},\ \bibinfo {pages} {470} (\bibinfo {year} {2020})}\BibitemShut {NoStop}%
\bibitem [{\citenamefont {Cvetkovic}\ \emph {et~al.}(2004)\citenamefont {Cvetkovic}, \citenamefont {Rowlinson},\ and\ \citenamefont {Simic}}]{line1-cvetkovic2004spectral}%
  \BibitemOpen
  \bibfield  {author} {\bibinfo {author} {\bibfnamefont {D.}~\bibnamefont {Cvetkovic}}, \bibinfo {author} {\bibfnamefont {P.}~\bibnamefont {Rowlinson}},\ and\ \bibinfo {author} {\bibfnamefont {S.}~\bibnamefont {Simic}},\ }\href@noop {} {\emph {\bibinfo {title} {Spectral generalizations of line graphs: On graphs with least eigenvalue-2}}},\ Vol.\ \bibinfo {volume} {314}\ (\bibinfo  {publisher} {Cambridge University Press},\ \bibinfo {year} {2004})\BibitemShut {NoStop}%
\bibitem [{\citenamefont {Chiu}\ \emph {et~al.}(2022)\citenamefont {Chiu}, \citenamefont {Carroll}, \citenamefont {Regnault},\ and\ \citenamefont {Houck}}]{line2-PhysRevResearch.4.023063}%
  \BibitemOpen
  \bibfield  {author} {\bibinfo {author} {\bibfnamefont {C.~S.}\ \bibnamefont {Chiu}}, \bibinfo {author} {\bibfnamefont {A.~N.}\ \bibnamefont {Carroll}}, \bibinfo {author} {\bibfnamefont {N.}~\bibnamefont {Regnault}},\ and\ \bibinfo {author} {\bibfnamefont {A.~A.}\ \bibnamefont {Houck}},\ }\bibfield  {title} {\bibinfo {title} {Line-graph-lattice crystal structures of stoichiometric materials},\ }\href {https://doi.org/10.1103/PhysRevResearch.4.023063} {\bibfield  {journal} {\bibinfo  {journal} {Phys. Rev. Res.}\ }\textbf {\bibinfo {volume} {4}},\ \bibinfo {pages} {023063} (\bibinfo {year} {2022})}\BibitemShut {NoStop}%
\bibitem [{\citenamefont {Koll{\'a}r}\ \emph {et~al.}(2020)\citenamefont {Koll{\'a}r}, \citenamefont {Fitzpatrick}, \citenamefont {Sarnak},\ and\ \citenamefont {Houck}}]{Kollar2020}%
  \BibitemOpen
  \bibfield  {author} {\bibinfo {author} {\bibfnamefont {A.~J.}\ \bibnamefont {Koll{\'a}r}}, \bibinfo {author} {\bibfnamefont {M.}~\bibnamefont {Fitzpatrick}}, \bibinfo {author} {\bibfnamefont {P.}~\bibnamefont {Sarnak}},\ and\ \bibinfo {author} {\bibfnamefont {A.~A.}\ \bibnamefont {Houck}},\ }\bibfield  {title} {\bibinfo {title} {Line-graph lattices: Euclidean and non-euclidean flat bands, and implementations in circuit quantum electrodynamics},\ }\href {https://doi.org/10.1007/s00220-019-03645-8} {\bibfield  {journal} {\bibinfo  {journal} {Communications in Mathematical Physics}\ }\textbf {\bibinfo {volume} {376}},\ \bibinfo {pages} {1909} (\bibinfo {year} {2020})}\BibitemShut {NoStop}%
\bibitem [{\citenamefont {Lieb}(1989)}]{Lieb1989}%
  \BibitemOpen
  \bibfield  {author} {\bibinfo {author} {\bibfnamefont {E.~H.}\ \bibnamefont {Lieb}},\ }\bibfield  {title} {\bibinfo {title} {Two theorems on the hubbard model},\ }\href {https://doi.org/10.1103/PhysRevLett.62.1201} {\bibfield  {journal} {\bibinfo  {journal} {Phys. Rev. Lett.}\ }\textbf {\bibinfo {volume} {62}},\ \bibinfo {pages} {1201} (\bibinfo {year} {1989})}\BibitemShut {NoStop}%
\bibitem [{\citenamefont {Sutherland}(1986)}]{Sutherland1986LocalizationTopology}%
  \BibitemOpen
  \bibfield  {author} {\bibinfo {author} {\bibfnamefont {B.}~\bibnamefont {Sutherland}},\ }\bibfield  {title} {\bibinfo {title} {{Localization of electronic wave functions due to local topology}},\ }\href {https://doi.org/10.1103/PhysRevB.34.5208} {\bibfield  {journal} {\bibinfo  {journal} {Physical Review B}\ }\textbf {\bibinfo {volume} {34}},\ \bibinfo {pages} {5208} (\bibinfo {year} {1986})}\BibitemShut {NoStop}%
\bibitem [{\citenamefont {Bercioux}\ \emph {et~al.}(2009)\citenamefont {Bercioux}, \citenamefont {Urban}, \citenamefont {Grabert},\ and\ \citenamefont {H{\"{a}}usler}}]{Bercioux2009MasslessLattice}%
  \BibitemOpen
  \bibfield  {author} {\bibinfo {author} {\bibfnamefont {D.}~\bibnamefont {Bercioux}}, \bibinfo {author} {\bibfnamefont {D.~F.}\ \bibnamefont {Urban}}, \bibinfo {author} {\bibfnamefont {H.}~\bibnamefont {Grabert}},\ and\ \bibinfo {author} {\bibfnamefont {W.}~\bibnamefont {H{\"{a}}usler}},\ }\bibfield  {title} {\bibinfo {title} {{Massless Dirac-Weyl fermions in a $\mathcal{T}_3$ optical lattice}},\ }\href {https://doi.org/10.1103/PhysRevA.80.063603} {\bibfield  {journal} {\bibinfo  {journal} {Physical Review A}\ }\textbf {\bibinfo {volume} {80}},\ \bibinfo {pages} {063603} (\bibinfo {year} {2009})}\BibitemShut {NoStop}%
\bibitem [{\citenamefont {Raoux}\ \emph {et~al.}(2014)\citenamefont {Raoux}, \citenamefont {Morigi}, \citenamefont {Fuchs}, \citenamefont {Pi{\'{e}}chon},\ and\ \citenamefont {Montambaux}}]{Raoux2014FromFermions}%
  \BibitemOpen
  \bibfield  {author} {\bibinfo {author} {\bibfnamefont {A.}~\bibnamefont {Raoux}}, \bibinfo {author} {\bibfnamefont {M.}~\bibnamefont {Morigi}}, \bibinfo {author} {\bibfnamefont {J.-N.}\ \bibnamefont {Fuchs}}, \bibinfo {author} {\bibfnamefont {F.}~\bibnamefont {Pi{\'{e}}chon}},\ and\ \bibinfo {author} {\bibfnamefont {G.}~\bibnamefont {Montambaux}},\ }\bibfield  {title} {\bibinfo {title} {{From Dia- to Paramagnetic Orbital Susceptibility of Massless Fermions}},\ }\href {https://doi.org/10.1103/PhysRevLett.112.026402} {\bibfield  {journal} {\bibinfo  {journal} {Physical Review Letters}\ }\textbf {\bibinfo {volume} {112}},\ \bibinfo {pages} {26402} (\bibinfo {year} {2014})}\BibitemShut {NoStop}%
\bibitem [{\citenamefont {Mojarro}\ \emph {et~al.}(2020{\natexlab{a}})\citenamefont {Mojarro}, \citenamefont {Ibarra-Sierra}, \citenamefont {Sandoval-Santana}, \citenamefont {Carrillo-Bastos},\ and\ \citenamefont {Naumis}}]{Mojarro20202}%
  \BibitemOpen
  \bibfield  {author} {\bibinfo {author} {\bibfnamefont {M.~A.}\ \bibnamefont {Mojarro}}, \bibinfo {author} {\bibfnamefont {V.~G.}\ \bibnamefont {Ibarra-Sierra}}, \bibinfo {author} {\bibfnamefont {J.~C.}\ \bibnamefont {Sandoval-Santana}}, \bibinfo {author} {\bibfnamefont {R.}~\bibnamefont {Carrillo-Bastos}},\ and\ \bibinfo {author} {\bibfnamefont {G.~G.}\ \bibnamefont {Naumis}},\ }\bibfield  {title} {\bibinfo {title} {Electron transitions for dirac hamiltonians with flat bands under electromagnetic radiation: Application to the $\alpha-\mathcal{T}_{3}$ graphene model},\ }\href {https://doi.org/10.1103/PhysRevB.101.165305} {\bibfield  {journal} {\bibinfo  {journal} {Phys. Rev. B}\ }\textbf {\bibinfo {volume} {101}},\ \bibinfo {pages} {165305} (\bibinfo {year} {2020}{\natexlab{a}})}\BibitemShut {NoStop}%
\bibitem [{\citenamefont {Oriekhov}\ \emph {et~al.}(2018)\citenamefont {Oriekhov}, \citenamefont {Gorbar},\ and\ \citenamefont {Gusynin}}]{Oriekhov2018ElectronicRibbon}%
  \BibitemOpen
  \bibfield  {author} {\bibinfo {author} {\bibfnamefont {D.~O.}\ \bibnamefont {Oriekhov}}, \bibinfo {author} {\bibfnamefont {E.~V.}\ \bibnamefont {Gorbar}},\ and\ \bibinfo {author} {\bibfnamefont {V.~P.}\ \bibnamefont {Gusynin}},\ }\bibfield  {title} {\bibinfo {title} {{Electronic states of pseudospin-1 fermions in dice lattice ribbon}},\ }\href {https://doi.org/10.1063/1.5078627} {\bibfield  {journal} {\bibinfo  {journal} {Low Temperature Physics}\ }\textbf {\bibinfo {volume} {44}},\ \bibinfo {pages} {1313} (\bibinfo {year} {2018})}\BibitemShut {NoStop}%
\bibitem [{\citenamefont {Tarnopolsky}\ \emph {et~al.}(2019)\citenamefont {Tarnopolsky}, \citenamefont {Kruchkov},\ and\ \citenamefont {Vishwanath}}]{Tarnopolsky2019OriginGraphene}%
  \BibitemOpen
  \bibfield  {author} {\bibinfo {author} {\bibfnamefont {G.}~\bibnamefont {Tarnopolsky}}, \bibinfo {author} {\bibfnamefont {A.~J.}\ \bibnamefont {Kruchkov}},\ and\ \bibinfo {author} {\bibfnamefont {A.}~\bibnamefont {Vishwanath}},\ }\bibfield  {title} {\bibinfo {title} {{Origin of Magic Angles in Twisted Bilayer Graphene}},\ }\href {https://doi.org/10.1103/PhysRevLett.122.106405} {\bibfield  {journal} {\bibinfo  {journal} {Physical Review Letters}\ }\textbf {\bibinfo {volume} {122}},\ \bibinfo {pages} {106405} (\bibinfo {year} {2019})}\BibitemShut {NoStop}%
\bibitem [{\citenamefont {Yu}\ and\ \citenamefont {Zhai}(2018)}]{Yu2018ChernLattices}%
  \BibitemOpen
  \bibfield  {author} {\bibinfo {author} {\bibfnamefont {H.~L.}\ \bibnamefont {Yu}}\ and\ \bibinfo {author} {\bibfnamefont {Z.~Y.}\ \bibnamefont {Zhai}},\ }\bibfield  {title} {\bibinfo {title} {{Chern number distribution and quantum phase transition in three-band lattices}},\ }\href {https://doi.org/10.1142/S0217984918501580} {\bibfield  {journal} {\bibinfo  {journal} {Modern Physics Letters B}\ }\textbf {\bibinfo {volume} {32}},\ \bibinfo {pages} {1850158} (\bibinfo {year} {2018})}\BibitemShut {NoStop}%
\bibitem [{\citenamefont {Yuan}\ and\ \citenamefont {Fu}(2018)}]{Yuan2018ModelBeyond}%
  \BibitemOpen
  \bibfield  {author} {\bibinfo {author} {\bibfnamefont {N.~F.~Q.}\ \bibnamefont {Yuan}}\ and\ \bibinfo {author} {\bibfnamefont {L.}~\bibnamefont {Fu}},\ }\bibfield  {title} {\bibinfo {title} {{Model for the metal-insulator transition in graphene superlattices and beyond}},\ }\href {https://doi.org/10.1103/PhysRevB.98.045103} {\bibfield  {journal} {\bibinfo  {journal} {Physical Review B}\ }\textbf {\bibinfo {volume} {98}},\ \bibinfo {pages} {045103} (\bibinfo {year} {2018})}\BibitemShut {NoStop}%
\bibitem [{\citenamefont {Illes}\ \emph {et~al.}(2015)\citenamefont {Illes}, \citenamefont {Carbotte},\ and\ \citenamefont {Nicol}}]{Illes2015HallModel}%
  \BibitemOpen
  \bibfield  {author} {\bibinfo {author} {\bibfnamefont {E.}~\bibnamefont {Illes}}, \bibinfo {author} {\bibfnamefont {J.~P.}\ \bibnamefont {Carbotte}},\ and\ \bibinfo {author} {\bibfnamefont {E.~J.}\ \bibnamefont {Nicol}},\ }\bibfield  {title} {\bibinfo {title} {{Hall quantization and optical conductivity evolution with variable Berry phase in the {$\alpha$}-$T_3$ model}},\ }\href {https://doi.org/10.1103/PhysRevB.92.245410} {\bibfield  {journal} {\bibinfo  {journal} {Physical Review B}\ }\textbf {\bibinfo {volume} {92}},\ \bibinfo {pages} {245410} (\bibinfo {year} {2015})}\BibitemShut {NoStop}%
\bibitem [{\citenamefont {Han}\ and\ \citenamefont {Lai}(2022)}]{HanOpticalResponseFlatBand2022}%
  \BibitemOpen
  \bibfield  {author} {\bibinfo {author} {\bibfnamefont {C.-D.}\ \bibnamefont {Han}}\ and\ \bibinfo {author} {\bibfnamefont {Y.-C.}\ \bibnamefont {Lai}},\ }\bibfield  {title} {\bibinfo {title} {Optical response of two-dimensional dirac materials with a flat band},\ }\href {https://doi.org/10.1103/PhysRevB.105.155405} {\bibfield  {journal} {\bibinfo  {journal} {Phys. Rev. B}\ }\textbf {\bibinfo {volume} {105}},\ \bibinfo {pages} {155405} (\bibinfo {year} {2022})}\BibitemShut {NoStop}%
\bibitem [{\citenamefont {Iurov}\ \emph {et~al.}(2023)\citenamefont {Iurov}, \citenamefont {Zhemchuzhna}, \citenamefont {Gumbs},\ and\ \citenamefont {Huang}}]{iurovoptical2023}%
  \BibitemOpen
  \bibfield  {author} {\bibinfo {author} {\bibfnamefont {A.}~\bibnamefont {Iurov}}, \bibinfo {author} {\bibfnamefont {L.}~\bibnamefont {Zhemchuzhna}}, \bibinfo {author} {\bibfnamefont {G.}~\bibnamefont {Gumbs}},\ and\ \bibinfo {author} {\bibfnamefont {D.}~\bibnamefont {Huang}},\ }\bibfield  {title} {\bibinfo {title} {Optical conductivity of gapped $\alpha-\mathcal{T}_{3}$ materials with a deformed flat band},\ }\href {https://doi.org/10.1103/PhysRevB.107.195137} {\bibfield  {journal} {\bibinfo  {journal} {Phys. Rev. B}\ }\textbf {\bibinfo {volume} {107}},\ \bibinfo {pages} {195137} (\bibinfo {year} {2023})}\BibitemShut {NoStop}%
\bibitem [{\citenamefont {Ye}\ \emph {et~al.}(2024)\citenamefont {Ye}, \citenamefont {Han},\ and\ \citenamefont {Lai}}]{ye2024optical}%
  \BibitemOpen
  \bibfield  {author} {\bibinfo {author} {\bibfnamefont {L.-L.}\ \bibnamefont {Ye}}, \bibinfo {author} {\bibfnamefont {C.-D.}\ \bibnamefont {Han}},\ and\ \bibinfo {author} {\bibfnamefont {Y.-C.}\ \bibnamefont {Lai}},\ }\bibfield  {title} {\bibinfo {title} {{Optical properties of two-dimensional Dirac–Weyl materials with a flatband}},\ }\href {https://doi.org/10.1063/5.0178936} {\bibfield  {journal} {\bibinfo  {journal} {Applied Physics Letters}\ }\textbf {\bibinfo {volume} {124}},\ \bibinfo {pages} {060501} (\bibinfo {year} {2024})}\BibitemShut {NoStop}%
\bibitem [{\citenamefont {Ponomarenko}\ \emph {et~al.}(2013)\citenamefont {Ponomarenko}, \citenamefont {Gorbachev}, \citenamefont {Yu}, \citenamefont {Elias}, \citenamefont {Jalil}, \citenamefont {Patel}, \citenamefont {Mishchenko}, \citenamefont {Mayorov}, \citenamefont {Woods}, \citenamefont {Wallbank} \emph {et~al.}}]{ponomarenko2013cloning}%
  \BibitemOpen
  \bibfield  {author} {\bibinfo {author} {\bibfnamefont {L.}~\bibnamefont {Ponomarenko}}, \bibinfo {author} {\bibfnamefont {R.}~\bibnamefont {Gorbachev}}, \bibinfo {author} {\bibfnamefont {G.}~\bibnamefont {Yu}}, \bibinfo {author} {\bibfnamefont {D.}~\bibnamefont {Elias}}, \bibinfo {author} {\bibfnamefont {R.}~\bibnamefont {Jalil}}, \bibinfo {author} {\bibfnamefont {A.}~\bibnamefont {Patel}}, \bibinfo {author} {\bibfnamefont {A.}~\bibnamefont {Mishchenko}}, \bibinfo {author} {\bibfnamefont {A.}~\bibnamefont {Mayorov}}, \bibinfo {author} {\bibfnamefont {C.}~\bibnamefont {Woods}}, \bibinfo {author} {\bibfnamefont {J.}~\bibnamefont {Wallbank}}, \emph {et~al.},\ }\bibfield  {title} {\bibinfo {title} {Cloning of dirac fermions in graphene superlattices},\ }\href {https://doi.org/10.1038/nature12187} {\bibfield  {journal} {\bibinfo  {journal} {Nature}\ }\textbf {\bibinfo {volume} {497}},\ \bibinfo {pages} {594} (\bibinfo {year} {2013})}\BibitemShut {NoStop}%
\bibitem [{\citenamefont {Hou}\ \emph {et~al.}(2007{\natexlab{a}})\citenamefont {Hou}, \citenamefont {Chamon},\ and\ \citenamefont {Mudry}}]{houelectron2007}%
  \BibitemOpen
  \bibfield  {author} {\bibinfo {author} {\bibfnamefont {C.-Y.}\ \bibnamefont {Hou}}, \bibinfo {author} {\bibfnamefont {C.}~\bibnamefont {Chamon}},\ and\ \bibinfo {author} {\bibfnamefont {C.}~\bibnamefont {Mudry}},\ }\bibfield  {title} {\bibinfo {title} {Electron fractionalization in two-dimensional graphenelike structures},\ }\href {https://doi.org/10.1103/PhysRevLett.98.186809} {\bibfield  {journal} {\bibinfo  {journal} {Phys. Rev. Lett.}\ }\textbf {\bibinfo {volume} {98}},\ \bibinfo {pages} {186809} (\bibinfo {year} {2007}{\natexlab{a}})}\BibitemShut {NoStop}%
\bibitem [{\citenamefont {Yankowitz}\ \emph {et~al.}(2012)\citenamefont {Yankowitz}, \citenamefont {Xue}, \citenamefont {Cormode}, \citenamefont {Sanchez-Yamagishi}, \citenamefont {Watanabe}, \citenamefont {Taniguchi}, \citenamefont {Jarillo-Herrero}, \citenamefont {Jacquod},\ and\ \citenamefont {LeRoy}}]{yankowitz2012emergence}%
  \BibitemOpen
  \bibfield  {author} {\bibinfo {author} {\bibfnamefont {M.}~\bibnamefont {Yankowitz}}, \bibinfo {author} {\bibfnamefont {J.}~\bibnamefont {Xue}}, \bibinfo {author} {\bibfnamefont {D.}~\bibnamefont {Cormode}}, \bibinfo {author} {\bibfnamefont {J.~D.}\ \bibnamefont {Sanchez-Yamagishi}}, \bibinfo {author} {\bibfnamefont {K.}~\bibnamefont {Watanabe}}, \bibinfo {author} {\bibfnamefont {T.}~\bibnamefont {Taniguchi}}, \bibinfo {author} {\bibfnamefont {P.}~\bibnamefont {Jarillo-Herrero}}, \bibinfo {author} {\bibfnamefont {P.}~\bibnamefont {Jacquod}},\ and\ \bibinfo {author} {\bibfnamefont {B.~J.}\ \bibnamefont {LeRoy}},\ }\bibfield  {title} {\bibinfo {title} {Emergence of superlattice dirac points in graphene on hexagonal boron nitride},\ }\href {https://doi.org/10.1038/nphys2272} {\bibfield  {journal} {\bibinfo  {journal} {Nature Physics}\ }\textbf {\bibinfo {volume} {8}},\ \bibinfo {pages} {382} (\bibinfo {year} {2012})}\BibitemShut {NoStop}%
\bibitem [{\citenamefont {Park}\ \emph {et~al.}(2008)\citenamefont {Park}, \citenamefont {Yang}, \citenamefont {Son}, \citenamefont {Cohen},\ and\ \citenamefont {Louie}}]{park2008anisotropic}%
  \BibitemOpen
  \bibfield  {author} {\bibinfo {author} {\bibfnamefont {C.-H.}\ \bibnamefont {Park}}, \bibinfo {author} {\bibfnamefont {L.}~\bibnamefont {Yang}}, \bibinfo {author} {\bibfnamefont {Y.-W.}\ \bibnamefont {Son}}, \bibinfo {author} {\bibfnamefont {M.~L.}\ \bibnamefont {Cohen}},\ and\ \bibinfo {author} {\bibfnamefont {S.~G.}\ \bibnamefont {Louie}},\ }\bibfield  {title} {\bibinfo {title} {Anisotropic behaviours of massless dirac fermions in graphene under periodic potentials},\ }\href {https://doi.org/10.1038/nphys890} {\bibfield  {journal} {\bibinfo  {journal} {Nature Physics}\ }\textbf {\bibinfo {volume} {4}},\ \bibinfo {pages} {213} (\bibinfo {year} {2008})}\BibitemShut {NoStop}%
\bibitem [{\citenamefont {Bao}\ \emph {et~al.}(2021)\citenamefont {Bao}, \citenamefont {Zhang}, \citenamefont {Zhang}, \citenamefont {Wu}, \citenamefont {Luo}, \citenamefont {Zhou}, \citenamefont {Li}, \citenamefont {Hou}, \citenamefont {Yao}, \citenamefont {Liu}, \citenamefont {Yu}, \citenamefont {Li}, \citenamefont {Duan}, \citenamefont {Yao}, \citenamefont {Wang},\ and\ \citenamefont {Zhou}}]{Bao2021ExperimentalGraphene}%
  \BibitemOpen
  \bibfield  {author} {\bibinfo {author} {\bibfnamefont {C.}~\bibnamefont {Bao}}, \bibinfo {author} {\bibfnamefont {H.}~\bibnamefont {Zhang}}, \bibinfo {author} {\bibfnamefont {T.}~\bibnamefont {Zhang}}, \bibinfo {author} {\bibfnamefont {X.}~\bibnamefont {Wu}}, \bibinfo {author} {\bibfnamefont {L.}~\bibnamefont {Luo}}, \bibinfo {author} {\bibfnamefont {S.}~\bibnamefont {Zhou}}, \bibinfo {author} {\bibfnamefont {Q.}~\bibnamefont {Li}}, \bibinfo {author} {\bibfnamefont {Y.}~\bibnamefont {Hou}}, \bibinfo {author} {\bibfnamefont {W.}~\bibnamefont {Yao}}, \bibinfo {author} {\bibfnamefont {L.}~\bibnamefont {Liu}}, \bibinfo {author} {\bibfnamefont {P.}~\bibnamefont {Yu}}, \bibinfo {author} {\bibfnamefont {J.}~\bibnamefont {Li}}, \bibinfo {author} {\bibfnamefont {W.}~\bibnamefont {Duan}}, \bibinfo {author} {\bibfnamefont {H.}~\bibnamefont {Yao}}, \bibinfo {author} {\bibfnamefont {Y.}~\bibnamefont {Wang}},\ and\ \bibinfo {author} {\bibfnamefont {S.}~\bibnamefont {Zhou}},\ }\bibfield  {title} {\bibinfo {title}
  {{Experimental Evidence of Chiral Symmetry Breaking in Kekul{\'{e}}-Ordered Graphene}},\ }\href {https://doi.org/10.1103/PhysRevLett.126.206804} {\bibfield  {journal} {\bibinfo  {journal} {Physical Review Letters}\ }\textbf {\bibinfo {volume} {126}},\ \bibinfo {pages} {206804} (\bibinfo {year} {2021})}\BibitemShut {NoStop}%
\bibitem [{\citenamefont {Gomes}\ \emph {et~al.}(2012)\citenamefont {Gomes}, \citenamefont {Mar}, \citenamefont {Ko}, \citenamefont {Guinea},\ and\ \citenamefont {Manoharan}}]{Gomes2012DesignerGraphene}%
  \BibitemOpen
  \bibfield  {author} {\bibinfo {author} {\bibfnamefont {K.~K.}\ \bibnamefont {Gomes}}, \bibinfo {author} {\bibfnamefont {W.}~\bibnamefont {Mar}}, \bibinfo {author} {\bibfnamefont {W.}~\bibnamefont {Ko}}, \bibinfo {author} {\bibfnamefont {F.}~\bibnamefont {Guinea}},\ and\ \bibinfo {author} {\bibfnamefont {H.~C.}\ \bibnamefont {Manoharan}},\ }\bibfield  {title} {\bibinfo {title} {{Designer Dirac fermions and topological phases in molecular graphene}},\ }\href {https://doi.org/10.1038/nature10941} {\bibfield  {journal} {\bibinfo  {journal} {Nature}\ }\textbf {\bibinfo {volume} {483}},\ \bibinfo {pages} {306} (\bibinfo {year} {2012})}\BibitemShut {NoStop}%
\bibitem [{\citenamefont {Guti{\'{e}}rrez}\ \emph {et~al.}(2016)\citenamefont {Guti{\'{e}}rrez}, \citenamefont {Kim}, \citenamefont {Brown}, \citenamefont {Schiros}, \citenamefont {Nordlund}, \citenamefont {Lochocki}, \citenamefont {Shen}, \citenamefont {Park},\ and\ \citenamefont {Pasupathy}}]{Gutierrez2016ImagingGraphene}%
  \BibitemOpen
  \bibfield  {author} {\bibinfo {author} {\bibfnamefont {C.}~\bibnamefont {Guti{\'{e}}rrez}}, \bibinfo {author} {\bibfnamefont {C.-J.}\ \bibnamefont {Kim}}, \bibinfo {author} {\bibfnamefont {L.}~\bibnamefont {Brown}}, \bibinfo {author} {\bibfnamefont {T.}~\bibnamefont {Schiros}}, \bibinfo {author} {\bibfnamefont {D.}~\bibnamefont {Nordlund}}, \bibinfo {author} {\bibfnamefont {E.}~\bibnamefont {Lochocki}}, \bibinfo {author} {\bibfnamefont {K.~M.}\ \bibnamefont {Shen}}, \bibinfo {author} {\bibfnamefont {J.}~\bibnamefont {Park}},\ and\ \bibinfo {author} {\bibfnamefont {A.~N.}\ \bibnamefont {Pasupathy}},\ }\bibfield  {title} {\bibinfo {title} {{Imaging chiral symmetry breaking from Kekul{\'{e}} bond order in graphene}},\ }\href {https://doi.org/10.1038/nphys3776} {\bibfield  {journal} {\bibinfo  {journal} {Nature Physics}\ }\textbf {\bibinfo {volume} {12}},\ \bibinfo {pages} {950} (\bibinfo {year} {2016})}\BibitemShut {NoStop}%
\bibitem [{\citenamefont {Gamayun}\ \emph {et~al.}(2018)\citenamefont {Gamayun}, \citenamefont {Ostroukh}, \citenamefont {Gnezdilov}, \citenamefont {Adagideli},\ and\ \citenamefont {Beenakker}}]{Gamayun2018Valley-momentumTexture}%
  \BibitemOpen
  \bibfield  {author} {\bibinfo {author} {\bibfnamefont {O.~V.}\ \bibnamefont {Gamayun}}, \bibinfo {author} {\bibfnamefont {V.~P.}\ \bibnamefont {Ostroukh}}, \bibinfo {author} {\bibfnamefont {N.~V.}\ \bibnamefont {Gnezdilov}}, \bibinfo {author} {\bibfnamefont {I.}~\bibnamefont {Adagideli}},\ and\ \bibinfo {author} {\bibfnamefont {C.~W.~J.}\ \bibnamefont {Beenakker}},\ }\bibfield  {title} {\bibinfo {title} {{Valley-momentum locking in a graphene superlattice with Y-shaped Kekul{\'{e}} bond texture}},\ }\href {https://doi.org/10.1088/1367-2630/aaa7e5} {\bibfield  {journal} {\bibinfo  {journal} {New Journal of Physics}\ }\textbf {\bibinfo {volume} {20}},\ \bibinfo {pages} {23016} (\bibinfo {year} {2018})}\BibitemShut {NoStop}%
\bibitem [{\citenamefont {Eom}\ and\ \citenamefont {Koo}(2020)}]{eomkekule2020}%
  \BibitemOpen
  \bibfield  {author} {\bibinfo {author} {\bibfnamefont {D.}~\bibnamefont {Eom}}\ and\ \bibinfo {author} {\bibfnamefont {J.-Y.}\ \bibnamefont {Koo}},\ }\bibfield  {title} {\bibinfo {title} {Direct measurement of strain-driven kekulé distortion in graphene and its electronic properties},\ }\href {https://doi.org/10.1039/D0NR03565C} {\bibfield  {journal} {\bibinfo  {journal} {Nanoscale}\ }\textbf {\bibinfo {volume} {12}},\ \bibinfo {pages} {19604} (\bibinfo {year} {2020})}\BibitemShut {NoStop}%
\bibitem [{\citenamefont {Qu}\ \emph {et~al.}(2022)\citenamefont {Qu}, \citenamefont {Nigge}, \citenamefont {Link}, \citenamefont {Levy}, \citenamefont {Michiardi}, \citenamefont {Spandar}, \citenamefont {Matthé}, \citenamefont {Schneider}, \citenamefont {Zhdanovich}, \citenamefont {Starke}, \citenamefont {Gutiérrez},\ and\ \citenamefont {Damascelli}}]{quubiqutuos2022}%
  \BibitemOpen
  \bibfield  {author} {\bibinfo {author} {\bibfnamefont {A.~C.}\ \bibnamefont {Qu}}, \bibinfo {author} {\bibfnamefont {P.}~\bibnamefont {Nigge}}, \bibinfo {author} {\bibfnamefont {S.}~\bibnamefont {Link}}, \bibinfo {author} {\bibfnamefont {G.}~\bibnamefont {Levy}}, \bibinfo {author} {\bibfnamefont {M.}~\bibnamefont {Michiardi}}, \bibinfo {author} {\bibfnamefont {P.~L.}\ \bibnamefont {Spandar}}, \bibinfo {author} {\bibfnamefont {T.}~\bibnamefont {Matthé}}, \bibinfo {author} {\bibfnamefont {M.}~\bibnamefont {Schneider}}, \bibinfo {author} {\bibfnamefont {S.}~\bibnamefont {Zhdanovich}}, \bibinfo {author} {\bibfnamefont {U.}~\bibnamefont {Starke}}, \bibinfo {author} {\bibfnamefont {C.}~\bibnamefont {Gutiérrez}},\ and\ \bibinfo {author} {\bibfnamefont {A.}~\bibnamefont {Damascelli}},\ }\bibfield  {title} {\bibinfo {title} {Ubiquitous defect-induced density wave instability in monolayer graphene},\ }\href {https://doi.org/10.1126/sciadv.abm5180} {\bibfield  {journal} {\bibinfo  {journal} {Science Advances}\
  }\textbf {\bibinfo {volume} {8}},\ \bibinfo {pages} {eabm5180} (\bibinfo {year} {2022})}\BibitemShut {NoStop}%
\bibitem [{\citenamefont {Cheianov}\ \emph {et~al.}(2009{\natexlab{a}})\citenamefont {Cheianov}, \citenamefont {Fal’ko}, \citenamefont {Sylju{\aa}sen},\ and\ \citenamefont {Altshuler}}]{Cheianov2009HiddenGraphene}%
  \BibitemOpen
  \bibfield  {author} {\bibinfo {author} {\bibfnamefont {V.~V.}\ \bibnamefont {Cheianov}}, \bibinfo {author} {\bibfnamefont {V.~I.}\ \bibnamefont {Fal’ko}}, \bibinfo {author} {\bibfnamefont {O.}~\bibnamefont {Sylju{\aa}sen}},\ and\ \bibinfo {author} {\bibfnamefont {B.~L.}\ \bibnamefont {Altshuler}},\ }\bibfield  {title} {\bibinfo {title} {{Hidden Kekul{\'{e}} ordering of adatoms on graphene}},\ }\href {https://doi.org/10.1016/j.ssc.2009.07.008} {\bibfield  {journal} {\bibinfo  {journal} {Solid State Communications}\ }\textbf {\bibinfo {volume} {149}},\ \bibinfo {pages} {1499} (\bibinfo {year} {2009}{\natexlab{a}})}\BibitemShut {NoStop}%
\bibitem [{\citenamefont {Cheianov}\ \emph {et~al.}(2009{\natexlab{b}})\citenamefont {Cheianov}, \citenamefont {Sylju{\aa}sen}, \citenamefont {Altshuler},\ and\ \citenamefont {Fal’ko}}]{Cheianov2009OrderedGraphene}%
  \BibitemOpen
  \bibfield  {author} {\bibinfo {author} {\bibfnamefont {V.~V.}\ \bibnamefont {Cheianov}}, \bibinfo {author} {\bibfnamefont {O.}~\bibnamefont {Sylju{\aa}sen}}, \bibinfo {author} {\bibfnamefont {B.~L.}\ \bibnamefont {Altshuler}},\ and\ \bibinfo {author} {\bibfnamefont {V.}~\bibnamefont {Fal’ko}},\ }\bibfield  {title} {\bibinfo {title} {{Ordered states of adatoms on graphene}},\ }\href {https://doi.org/10.1103/PhysRevB.80.233409} {\bibfield  {journal} {\bibinfo  {journal} {Physical Review B}\ }\textbf {\bibinfo {volume} {80}},\ \bibinfo {pages} {233409} (\bibinfo {year} {2009}{\natexlab{b}})}\BibitemShut {NoStop}%
\bibitem [{\citenamefont {Farjam}\ and\ \citenamefont {Rafii-Tabar}(2009)}]{Farjam2009EnergyCalculations}%
  \BibitemOpen
  \bibfield  {author} {\bibinfo {author} {\bibfnamefont {M.}~\bibnamefont {Farjam}}\ and\ \bibinfo {author} {\bibfnamefont {H.}~\bibnamefont {Rafii-Tabar}},\ }\bibfield  {title} {\bibinfo {title} {{Energy gap opening in submonolayer lithium on graphene: Local density functional and tight-binding calculations}},\ }\href {https://doi.org/10.1103/PhysRevB.79.045417} {\bibfield  {journal} {\bibinfo  {journal} {Physical Review B}\ }\textbf {\bibinfo {volume} {79}},\ \bibinfo {pages} {045417} (\bibinfo {year} {2009})}\BibitemShut {NoStop}%
\bibitem [{\citenamefont {Sugawara}\ \emph {et~al.}(2011)\citenamefont {Sugawara}, \citenamefont {Kanetani}, \citenamefont {Sato},\ and\ \citenamefont {Takahashi}}]{Sugawara2011FabricationGraphene}%
  \BibitemOpen
  \bibfield  {author} {\bibinfo {author} {\bibfnamefont {K.}~\bibnamefont {Sugawara}}, \bibinfo {author} {\bibfnamefont {K.}~\bibnamefont {Kanetani}}, \bibinfo {author} {\bibfnamefont {T.}~\bibnamefont {Sato}},\ and\ \bibinfo {author} {\bibfnamefont {T.}~\bibnamefont {Takahashi}},\ }\bibfield  {title} {\bibinfo {title} {Fabrication of li-intercalated bilayer graphene},\ }\href {https://doi.org/10.1063/1.3582814} {\bibfield  {journal} {\bibinfo  {journal} {AIP Advances}\ }\textbf {\bibinfo {volume} {1}},\ \bibinfo {pages} {22103} (\bibinfo {year} {2011})}\BibitemShut {NoStop}%
\bibitem [{\citenamefont {Kanetani}\ \emph {et~al.}(2012)\citenamefont {Kanetani}, \citenamefont {Sugawara}, \citenamefont {Sato}, \citenamefont {Shimizu}, \citenamefont {Iwaya}, \citenamefont {Hitosugi},\ and\ \citenamefont {Takahashi}}]{Kanetani2012CaCa}%
  \BibitemOpen
  \bibfield  {author} {\bibinfo {author} {\bibfnamefont {K.}~\bibnamefont {Kanetani}}, \bibinfo {author} {\bibfnamefont {K.}~\bibnamefont {Sugawara}}, \bibinfo {author} {\bibfnamefont {T.}~\bibnamefont {Sato}}, \bibinfo {author} {\bibfnamefont {R.}~\bibnamefont {Shimizu}}, \bibinfo {author} {\bibfnamefont {K.}~\bibnamefont {Iwaya}}, \bibinfo {author} {\bibfnamefont {T.}~\bibnamefont {Hitosugi}},\ and\ \bibinfo {author} {\bibfnamefont {T.}~\bibnamefont {Takahashi}},\ }\bibfield  {title} {\bibinfo {title} {{Ca intercalated bilayer graphene as a thinnest limit of superconducting C$_6$ Ca}},\ }\href {https://doi.org/10.1073/pnas.1208889109} {\bibfield  {journal} {\bibinfo  {journal} {Proceedings of the National Academy of Sciences}\ }\textbf {\bibinfo {volume} {109}},\ \bibinfo {pages} {19610} (\bibinfo {year} {2012})}\BibitemShut {NoStop}%
\bibitem [{\citenamefont {Chamon}(2000)}]{Chamon2000SolitonsNanotubes}%
  \BibitemOpen
  \bibfield  {author} {\bibinfo {author} {\bibfnamefont {C.}~\bibnamefont {Chamon}},\ }\bibfield  {title} {\bibinfo {title} {{Solitons in carbon nanotubes}},\ }\href {https://doi.org/10.1103/PhysRevB.62.2806} {\bibfield  {journal} {\bibinfo  {journal} {Physical Review B}\ }\textbf {\bibinfo {volume} {62}},\ \bibinfo {pages} {2806} (\bibinfo {year} {2000})}\BibitemShut {NoStop}%
\bibitem [{\citenamefont {Classen}\ \emph {et~al.}(2014)\citenamefont {Classen}, \citenamefont {Scherer},\ and\ \citenamefont {Honerkamp}}]{Classen2014InstabilitiesInteractions}%
  \BibitemOpen
  \bibfield  {author} {\bibinfo {author} {\bibfnamefont {L.}~\bibnamefont {Classen}}, \bibinfo {author} {\bibfnamefont {M.~M.}\ \bibnamefont {Scherer}},\ and\ \bibinfo {author} {\bibfnamefont {C.}~\bibnamefont {Honerkamp}},\ }\bibfield  {title} {\bibinfo {title} {Instabilities on graphene's honeycomb lattice with electron-phonon interactions},\ }\href {https://doi.org/10.1103/PhysRevB.90.035122} {\bibfield  {journal} {\bibinfo  {journal} {Phys. Rev. B}\ }\textbf {\bibinfo {volume} {90}},\ \bibinfo {pages} {035122} (\bibinfo {year} {2014})}\BibitemShut {NoStop}%
\bibitem [{\citenamefont {Weeks}\ and\ \citenamefont {Franz}(2010)}]{Weeks2010Interaction-drivenSemimetal}%
  \BibitemOpen
  \bibfield  {author} {\bibinfo {author} {\bibfnamefont {C.}~\bibnamefont {Weeks}}\ and\ \bibinfo {author} {\bibfnamefont {M.}~\bibnamefont {Franz}},\ }\bibfield  {title} {\bibinfo {title} {{Interaction-driven instabilities of a Dirac semimetal}},\ }\href {https://doi.org/10.1103/PhysRevB.81.085105} {\bibfield  {journal} {\bibinfo  {journal} {Physical Review B}\ }\textbf {\bibinfo {volume} {81}},\ \bibinfo {pages} {85105} (\bibinfo {year} {2010})}\BibitemShut {NoStop}%
\bibitem [{\citenamefont {Hou}\ \emph {et~al.}(2007{\natexlab{b}})\citenamefont {Hou}, \citenamefont {Chamon},\ and\ \citenamefont {Mudry}}]{Hou2007ElectronStructures}%
  \BibitemOpen
  \bibfield  {author} {\bibinfo {author} {\bibfnamefont {C.-Y.}\ \bibnamefont {Hou}}, \bibinfo {author} {\bibfnamefont {C.}~\bibnamefont {Chamon}},\ and\ \bibinfo {author} {\bibfnamefont {C.}~\bibnamefont {Mudry}},\ }\bibfield  {title} {\bibinfo {title} {{Electron Fractionalization in Two-Dimensional Graphenelike Structures}},\ }\href {https://doi.org/10.1103/PhysRevLett.98.186809} {\bibfield  {journal} {\bibinfo  {journal} {Physical Review Letters}\ }\textbf {\bibinfo {volume} {98}},\ \bibinfo {pages} {186809} (\bibinfo {year} {2007}{\natexlab{b}})}\BibitemShut {NoStop}%
\bibitem [{\citenamefont {Marianetti}\ and\ \citenamefont {Yevick}(2010)}]{Marianetti2010FailureTension}%
  \BibitemOpen
  \bibfield  {author} {\bibinfo {author} {\bibfnamefont {C.~A.}\ \bibnamefont {Marianetti}}\ and\ \bibinfo {author} {\bibfnamefont {H.~G.}\ \bibnamefont {Yevick}},\ }\bibfield  {title} {\bibinfo {title} {{Failure Mechanisms of Graphene under Tension}},\ }\href {https://doi.org/10.1103/PhysRevLett.105.245502} {\bibfield  {journal} {\bibinfo  {journal} {Physical Review Letters}\ }\textbf {\bibinfo {volume} {105}},\ \bibinfo {pages} {245502} (\bibinfo {year} {2010})}\BibitemShut {NoStop}%
\bibitem [{\citenamefont {Lee}\ \emph {et~al.}(2011)\citenamefont {Lee}, \citenamefont {Chung}, \citenamefont {Heo}, \citenamefont {Yang}, \citenamefont {Shin}, \citenamefont {Chung},\ and\ \citenamefont {Seo}}]{Lee2011BandGraphene}%
  \BibitemOpen
  \bibfield  {author} {\bibinfo {author} {\bibfnamefont {S.-H.}\ \bibnamefont {Lee}}, \bibinfo {author} {\bibfnamefont {H.-J.}\ \bibnamefont {Chung}}, \bibinfo {author} {\bibfnamefont {J.}~\bibnamefont {Heo}}, \bibinfo {author} {\bibfnamefont {H.}~\bibnamefont {Yang}}, \bibinfo {author} {\bibfnamefont {J.}~\bibnamefont {Shin}}, \bibinfo {author} {\bibfnamefont {U.-I.}\ \bibnamefont {Chung}},\ and\ \bibinfo {author} {\bibfnamefont {S.}~\bibnamefont {Seo}},\ }\bibfield  {title} {\bibinfo {title} {{Band Gap Opening by Two-Dimensional Manifestation of Peierls Instability in Graphene}},\ }\href {https://doi.org/10.1021/nn1035894} {\bibfield  {journal} {\bibinfo  {journal} {ACS Nano}\ }\textbf {\bibinfo {volume} {5}},\ \bibinfo {pages} {2964} (\bibinfo {year} {2011})}\BibitemShut {NoStop}%
\bibitem [{\citenamefont {Giovannetti}\ \emph {et~al.}(2015{\natexlab{a}})\citenamefont {Giovannetti}, \citenamefont {Capone}, \citenamefont {Van Den~Brink},\ and\ \citenamefont {Ortix}}]{Giovannetti2015KekuleBilayer}%
  \BibitemOpen
  \bibfield  {author} {\bibinfo {author} {\bibfnamefont {G.}~\bibnamefont {Giovannetti}}, \bibinfo {author} {\bibfnamefont {M.}~\bibnamefont {Capone}}, \bibinfo {author} {\bibfnamefont {J.}~\bibnamefont {Van Den~Brink}},\ and\ \bibinfo {author} {\bibfnamefont {C.}~\bibnamefont {Ortix}},\ }\bibfield  {title} {\bibinfo {title} {{Kekul{\'{e}} textures, pseudospin-one Dirac cones, and quadratic band crossings in a graphene-hexagonal indium chalcogenide bilayer}},\ }\href {https://doi.org/10.1103/PhysRevB.91.121417} {\bibfield  {journal} {\bibinfo  {journal} {Physical Review B}\ }\textbf {\bibinfo {volume} {91}},\ \bibinfo {pages} {121417} (\bibinfo {year} {2015}{\natexlab{a}})}\BibitemShut {NoStop}%
\bibitem [{\citenamefont {Im}\ \emph {et~al.}(2023)\citenamefont {Im}, \citenamefont {Im}, \citenamefont {Kim}, \citenamefont {Lee}, \citenamefont {Hwang}, \citenamefont {Mo},\ and\ \citenamefont {Hwang}}]{Im2023ModifiedHeterostructure}%
  \BibitemOpen
  \bibfield  {author} {\bibinfo {author} {\bibfnamefont {S.}~\bibnamefont {Im}}, \bibinfo {author} {\bibfnamefont {H.}~\bibnamefont {Im}}, \bibinfo {author} {\bibfnamefont {K.}~\bibnamefont {Kim}}, \bibinfo {author} {\bibfnamefont {J.}~\bibnamefont {Lee}}, \bibinfo {author} {\bibfnamefont {J.}~\bibnamefont {Hwang}}, \bibinfo {author} {\bibfnamefont {S.}~\bibnamefont {Mo}},\ and\ \bibinfo {author} {\bibfnamefont {C.}~\bibnamefont {Hwang}},\ }\bibfield  {title} {\bibinfo {title} {{Modified Dirac Fermions in the Crystalline Xenon and Graphene Moir{\'{e}} Heterostructure}},\ }\href {https://doi.org/10.1002/apxr.202200091} {\bibfield  {journal} {\bibinfo  {journal} {Advanced Physics Research}\ }\textbf {\bibinfo {volume} {2}},\ \bibinfo {pages} {2200091} (\bibinfo {year} {2023})}\BibitemShut {NoStop}%
\bibitem [{\citenamefont {Ye}\ \emph {et~al.}(2023)\citenamefont {Ye}, \citenamefont {Qian}, \citenamefont {Zhang}, \citenamefont {Wang}, \citenamefont {Xiao},\ and\ \citenamefont {Cao}}]{Ye2023KekuleSuperlattices}%
  \BibitemOpen
  \bibfield  {author} {\bibinfo {author} {\bibfnamefont {Y.}~\bibnamefont {Ye}}, \bibinfo {author} {\bibfnamefont {J.}~\bibnamefont {Qian}}, \bibinfo {author} {\bibfnamefont {X.-W.}\ \bibnamefont {Zhang}}, \bibinfo {author} {\bibfnamefont {C.}~\bibnamefont {Wang}}, \bibinfo {author} {\bibfnamefont {D.}~\bibnamefont {Xiao}},\ and\ \bibinfo {author} {\bibfnamefont {T.}~\bibnamefont {Cao}},\ }\bibfield  {title} {\bibinfo {title} {{Kekul{\'{e}} Moir{\'{e}} Superlattices}},\ }\href {https://doi.org/10.1021/acs.nanolett.3c01550} {\bibfield  {journal} {\bibinfo  {journal} {Nano Letters}\ }\textbf {\bibinfo {volume} {23}},\ \bibinfo {pages} {6536} (\bibinfo {year} {2023})}\BibitemShut {NoStop}%
\bibitem [{\citenamefont {Herrera}\ and\ \citenamefont {Naumis}(2020{\natexlab{a}})}]{HerreraElectronicOpticalKekY}%
  \BibitemOpen
  \bibfield  {author} {\bibinfo {author} {\bibfnamefont {S.~A.}\ \bibnamefont {Herrera}}\ and\ \bibinfo {author} {\bibfnamefont {G.~G.}\ \bibnamefont {Naumis}},\ }\bibfield  {title} {\bibinfo {title} {Electronic and optical conductivity of kekul\'e-patterned graphene: Intravalley and intervalley transport},\ }\href {https://doi.org/10.1103/PhysRevB.101.205413} {\bibfield  {journal} {\bibinfo  {journal} {Phys. Rev. B}\ }\textbf {\bibinfo {volume} {101}},\ \bibinfo {pages} {205413} (\bibinfo {year} {2020}{\natexlab{a}})}\BibitemShut {NoStop}%
\bibitem [{\citenamefont {Herrera}\ and\ \citenamefont {Naumis}(2020{\natexlab{b}})}]{HerreraDynamicPlasmonsKekule2020}%
  \BibitemOpen
  \bibfield  {author} {\bibinfo {author} {\bibfnamefont {S.~A.}\ \bibnamefont {Herrera}}\ and\ \bibinfo {author} {\bibfnamefont {G.~G.}\ \bibnamefont {Naumis}},\ }\bibfield  {title} {\bibinfo {title} {Dynamic polarization and plasmons in kekul\'e-patterned graphene: Signatures of broken valley degeneracy},\ }\href {https://doi.org/10.1103/PhysRevB.102.205429} {\bibfield  {journal} {\bibinfo  {journal} {Phys. Rev. B}\ }\textbf {\bibinfo {volume} {102}},\ \bibinfo {pages} {205429} (\bibinfo {year} {2020}{\natexlab{b}})}\BibitemShut {NoStop}%
\bibitem [{\citenamefont {Santacruz}\ \emph {et~al.}(2022)\citenamefont {Santacruz}, \citenamefont {Iglesias}, \citenamefont {Carrillo-Bastos},\ and\ \citenamefont {Mireles}}]{ValleydrivenSantacruz2022}%
  \BibitemOpen
  \bibfield  {author} {\bibinfo {author} {\bibfnamefont {A.}~\bibnamefont {Santacruz}}, \bibinfo {author} {\bibfnamefont {P.~E.}\ \bibnamefont {Iglesias}}, \bibinfo {author} {\bibfnamefont {R.}~\bibnamefont {Carrillo-Bastos}},\ and\ \bibinfo {author} {\bibfnamefont {F.}~\bibnamefont {Mireles}},\ }\bibfield  {title} {\bibinfo {title} {Valley-driven zitterbewegung in kekul\'e-distorted graphene},\ }\href {https://doi.org/10.1103/PhysRevB.105.205405} {\bibfield  {journal} {\bibinfo  {journal} {Phys. Rev. B}\ }\textbf {\bibinfo {volume} {105}},\ \bibinfo {pages} {205405} (\bibinfo {year} {2022})}\BibitemShut {NoStop}%
\bibitem [{\citenamefont {Andrade}\ \emph {et~al.}(2022)\citenamefont {Andrade}, \citenamefont {Carrillo-Bastos}, \citenamefont {Asmar},\ and\ \citenamefont {Naumis}}]{AndradeKekuléValleyBire2022}%
  \BibitemOpen
  \bibfield  {author} {\bibinfo {author} {\bibfnamefont {E.}~\bibnamefont {Andrade}}, \bibinfo {author} {\bibfnamefont {R.}~\bibnamefont {Carrillo-Bastos}}, \bibinfo {author} {\bibfnamefont {M.~M.}\ \bibnamefont {Asmar}},\ and\ \bibinfo {author} {\bibfnamefont {G.~G.}\ \bibnamefont {Naumis}},\ }\bibfield  {title} {\bibinfo {title} {Kekul\'e-induced valley birefringence and skew scattering in graphene},\ }\href {https://doi.org/10.1103/PhysRevB.106.195413} {\bibfield  {journal} {\bibinfo  {journal} {Phys. Rev. B}\ }\textbf {\bibinfo {volume} {106}},\ \bibinfo {pages} {195413} (\bibinfo {year} {2022})}\BibitemShut {NoStop}%
\bibitem [{\citenamefont {Herrera}\ and\ \citenamefont {Naumis}(2021)}]{HerreraOptoelectronic2021}%
  \BibitemOpen
  \bibfield  {author} {\bibinfo {author} {\bibfnamefont {S.~A.}\ \bibnamefont {Herrera}}\ and\ \bibinfo {author} {\bibfnamefont {G.~G.}\ \bibnamefont {Naumis}},\ }\bibfield  {title} {\bibinfo {title} {Optoelectronic fingerprints of interference between different charge carriers and band flattening in graphene superlattices},\ }\href {https://doi.org/10.1103/PhysRevB.104.115424} {\bibfield  {journal} {\bibinfo  {journal} {Phys. Rev. B}\ }\textbf {\bibinfo {volume} {104}},\ \bibinfo {pages} {115424} (\bibinfo {year} {2021})}\BibitemShut {NoStop}%
\bibitem [{\citenamefont {Mohammadi}(2021)}]{MohammadiOpticalKek2021}%
  \BibitemOpen
  \bibfield  {author} {\bibinfo {author} {\bibfnamefont {Y.}~\bibnamefont {Mohammadi}},\ }\bibfield  {title} {\bibinfo {title} {Magneto-optical conductivity of graphene: Signatures of a uniform y-shaped kekule lattice distortion},\ }\href {https://doi.org/10.1149/2162-8777/ac08d5} {\bibfield  {journal} {\bibinfo  {journal} {ECS Journal of Solid State Science and Technology}\ }\textbf {\bibinfo {volume} {10}},\ \bibinfo {pages} {061011} (\bibinfo {year} {2021})}\BibitemShut {NoStop}%
\bibitem [{\citenamefont {Mohammadi}(2022)}]{MohammadiElectronicKek2022}%
  \BibitemOpen
  \bibfield  {author} {\bibinfo {author} {\bibfnamefont {Y.}~\bibnamefont {Mohammadi}},\ }\bibfield  {title} {\bibinfo {title} {Electronic spectrum and optical properties of y-shaped kekulé-patterned graphene: Band nesting resonance as an optical signature},\ }\href {https://doi.org/10.1149/2162-8777/aca99b} {\bibfield  {journal} {\bibinfo  {journal} {ECS Journal of Solid State Science and Technology}\ }\textbf {\bibinfo {volume} {11}},\ \bibinfo {pages} {121004} (\bibinfo {year} {2022})}\BibitemShut {NoStop}%
\bibitem [{\citenamefont {Mojarro}\ \emph {et~al.}(2020{\natexlab{b}})\citenamefont {Mojarro}, \citenamefont {Ibarra-Sierra}, \citenamefont {Sandoval-Santana}, \citenamefont {Carrillo-Bastos},\ and\ \citenamefont {Naumis}}]{Mojarro2020}%
  \BibitemOpen
  \bibfield  {author} {\bibinfo {author} {\bibfnamefont {M.~A.}\ \bibnamefont {Mojarro}}, \bibinfo {author} {\bibfnamefont {V.~G.}\ \bibnamefont {Ibarra-Sierra}}, \bibinfo {author} {\bibfnamefont {J.~C.}\ \bibnamefont {Sandoval-Santana}}, \bibinfo {author} {\bibfnamefont {R.}~\bibnamefont {Carrillo-Bastos}},\ and\ \bibinfo {author} {\bibfnamefont {G.~G.}\ \bibnamefont {Naumis}},\ }\bibfield  {title} {\bibinfo {title} {Dynamical floquet spectrum of kekul\'e-distorted graphene under normal incidence of electromagnetic radiation},\ }\href {https://doi.org/10.1103/PhysRevB.102.165301} {\bibfield  {journal} {\bibinfo  {journal} {Phys. Rev. B}\ }\textbf {\bibinfo {volume} {102}},\ \bibinfo {pages} {165301} (\bibinfo {year} {2020}{\natexlab{b}})}\BibitemShut {NoStop}%
\bibitem [{\citenamefont {Tajkov}\ \emph {et~al.}(2020)\citenamefont {Tajkov}, \citenamefont {Koltai}, \citenamefont {Cserti},\ and\ \citenamefont {Oroszl\'any}}]{tajkov2020}%
  \BibitemOpen
  \bibfield  {author} {\bibinfo {author} {\bibfnamefont {Z.}~\bibnamefont {Tajkov}}, \bibinfo {author} {\bibfnamefont {J.}~\bibnamefont {Koltai}}, \bibinfo {author} {\bibfnamefont {J.}~\bibnamefont {Cserti}},\ and\ \bibinfo {author} {\bibfnamefont {L.}~\bibnamefont {Oroszl\'any}},\ }\bibfield  {title} {\bibinfo {title} {Competition of topological and topologically trivial phases in patterned graphene based heterostructures},\ }\href {https://doi.org/10.1103/PhysRevB.101.235146} {\bibfield  {journal} {\bibinfo  {journal} {Phys. Rev. B}\ }\textbf {\bibinfo {volume} {101}},\ \bibinfo {pages} {235146} (\bibinfo {year} {2020})}\BibitemShut {NoStop}%
\bibitem [{\citenamefont {Xu}\ \emph {et~al.}(2018)\citenamefont {Xu}, \citenamefont {Law},\ and\ \citenamefont {Lee}}]{XuKekule2018}%
  \BibitemOpen
  \bibfield  {author} {\bibinfo {author} {\bibfnamefont {X.~Y.}\ \bibnamefont {Xu}}, \bibinfo {author} {\bibfnamefont {K.~T.}\ \bibnamefont {Law}},\ and\ \bibinfo {author} {\bibfnamefont {P.~A.}\ \bibnamefont {Lee}},\ }\bibfield  {title} {\bibinfo {title} {Kekul\'e valence bond order in an extended hubbard model on the honeycomb lattice with possible applications to twisted bilayer graphene},\ }\href {https://doi.org/10.1103/PhysRevB.98.121406} {\bibfield  {journal} {\bibinfo  {journal} {Phys. Rev. B}\ }\textbf {\bibinfo {volume} {98}},\ \bibinfo {pages} {121406} (\bibinfo {year} {2018})}\BibitemShut {NoStop}%
\bibitem [{\citenamefont {Roy}\ and\ \citenamefont {Herbut}(2010)}]{RoyKekule2010}%
  \BibitemOpen
  \bibfield  {author} {\bibinfo {author} {\bibfnamefont {B.}~\bibnamefont {Roy}}\ and\ \bibinfo {author} {\bibfnamefont {I.~F.}\ \bibnamefont {Herbut}},\ }\bibfield  {title} {\bibinfo {title} {Unconventional superconductivity on honeycomb lattice: Theory of {K}ekule order parameter},\ }\href {https://doi.org/10.1103/PhysRevB.82.035429} {\bibfield  {journal} {\bibinfo  {journal} {Phys. Rev. B}\ }\textbf {\bibinfo {volume} {82}},\ \bibinfo {pages} {035429} (\bibinfo {year} {2010})}\BibitemShut {NoStop}%
\bibitem [{\citenamefont {Po}\ \emph {et~al.}(2018)\citenamefont {Po}, \citenamefont {Zou}, \citenamefont {Vishwanath},\ and\ \citenamefont {Senthil}}]{Po2018}%
  \BibitemOpen
  \bibfield  {author} {\bibinfo {author} {\bibfnamefont {H.~C.}\ \bibnamefont {Po}}, \bibinfo {author} {\bibfnamefont {L.}~\bibnamefont {Zou}}, \bibinfo {author} {\bibfnamefont {A.}~\bibnamefont {Vishwanath}},\ and\ \bibinfo {author} {\bibfnamefont {T.}~\bibnamefont {Senthil}},\ }\bibfield  {title} {\bibinfo {title} {Origin of mott insulating behavior and superconductivity in twisted bilayer graphene},\ }\href {https://doi.org/10.1103/PhysRevX.8.031089} {\bibfield  {journal} {\bibinfo  {journal} {Phys. Rev. X}\ }\textbf {\bibinfo {volume} {8}},\ \bibinfo {pages} {031089} (\bibinfo {year} {2018})}\BibitemShut {NoStop}%
\bibitem [{\citenamefont {Da~Liao}\ \emph {et~al.}(2019)\citenamefont {Da~Liao}, \citenamefont {Meng},\ and\ \citenamefont {Xu}}]{Da2019}%
  \BibitemOpen
  \bibfield  {author} {\bibinfo {author} {\bibfnamefont {Y.}~\bibnamefont {Da~Liao}}, \bibinfo {author} {\bibfnamefont {Z.~Y.}\ \bibnamefont {Meng}},\ and\ \bibinfo {author} {\bibfnamefont {X.~Y.}\ \bibnamefont {Xu}},\ }\bibfield  {title} {\bibinfo {title} {Valence bond orders at charge neutrality in a possible two-orbital extended hubbard model for twisted bilayer graphene},\ }\href {https://doi.org/10.1103/PhysRevLett.123.157601} {\bibfield  {journal} {\bibinfo  {journal} {Phys. Rev. Lett.}\ }\textbf {\bibinfo {volume} {123}},\ \bibinfo {pages} {157601} (\bibinfo {year} {2019})}\BibitemShut {NoStop}%
\bibitem [{\citenamefont {Huang}\ \emph {et~al.}(2020)\citenamefont {Huang}, \citenamefont {Huang},\ and\ \citenamefont {Lee}}]{Huangslave2020}%
  \BibitemOpen
  \bibfield  {author} {\bibinfo {author} {\bibfnamefont {S.-M.}\ \bibnamefont {Huang}}, \bibinfo {author} {\bibfnamefont {Y.-P.}\ \bibnamefont {Huang}},\ and\ \bibinfo {author} {\bibfnamefont {T.-K.}\ \bibnamefont {Lee}},\ }\bibfield  {title} {\bibinfo {title} {Slave-rotor theory on magic-angle twisted bilayer graphene},\ }\href {https://doi.org/10.1103/PhysRevB.101.235140} {\bibfield  {journal} {\bibinfo  {journal} {Phys. Rev. B}\ }\textbf {\bibinfo {volume} {101}},\ \bibinfo {pages} {235140} (\bibinfo {year} {2020})}\BibitemShut {NoStop}%
\bibitem [{\citenamefont {Blason}\ and\ \citenamefont {Fabrizio}(2022)}]{localkekule2022}%
  \BibitemOpen
  \bibfield  {author} {\bibinfo {author} {\bibfnamefont {A.}~\bibnamefont {Blason}}\ and\ \bibinfo {author} {\bibfnamefont {M.}~\bibnamefont {Fabrizio}},\ }\bibfield  {title} {\bibinfo {title} {Local kekul\'e distortion turns twisted bilayer graphene into topological mott insulators and superconductors},\ }\href {https://doi.org/10.1103/PhysRevB.106.235112} {\bibfield  {journal} {\bibinfo  {journal} {Phys. Rev. B}\ }\textbf {\bibinfo {volume} {106}},\ \bibinfo {pages} {235112} (\bibinfo {year} {2022})}\BibitemShut {NoStop}%
\bibitem [{\citenamefont {Wang}\ \emph {et~al.}(2014)\citenamefont {Wang}, \citenamefont {Xu},\ and\ \citenamefont {Zhang}}]{Wang2DSupeconductivity2014}%
  \BibitemOpen
  \bibfield  {author} {\bibinfo {author} {\bibfnamefont {J.}~\bibnamefont {Wang}}, \bibinfo {author} {\bibfnamefont {Y.}~\bibnamefont {Xu}},\ and\ \bibinfo {author} {\bibfnamefont {S.-C.}\ \bibnamefont {Zhang}},\ }\bibfield  {title} {\bibinfo {title} {Two-dimensional time-reversal-invariant topological superconductivity in a doped quantum spin-hall insulator},\ }\href {https://doi.org/10.1103/PhysRevB.90.054503} {\bibfield  {journal} {\bibinfo  {journal} {Phys. Rev. B}\ }\textbf {\bibinfo {volume} {90}},\ \bibinfo {pages} {054503} (\bibinfo {year} {2014})}\BibitemShut {NoStop}%
\bibitem [{\citenamefont {Gonz\'alez-\'Arraga}\ \emph {et~al.}(2018)\citenamefont {Gonz\'alez-\'Arraga}, \citenamefont {Guinea},\ and\ \citenamefont {San-Jose}}]{modulationarraga2018}%
  \BibitemOpen
  \bibfield  {author} {\bibinfo {author} {\bibfnamefont {L.}~\bibnamefont {Gonz\'alez-\'Arraga}}, \bibinfo {author} {\bibfnamefont {F.}~\bibnamefont {Guinea}},\ and\ \bibinfo {author} {\bibfnamefont {P.}~\bibnamefont {San-Jose}},\ }\bibfield  {title} {\bibinfo {title} {Modulation of kekul\'e adatom ordering due to strain in graphene},\ }\href {https://doi.org/10.1103/PhysRevB.97.165430} {\bibfield  {journal} {\bibinfo  {journal} {Phys. Rev. B}\ }\textbf {\bibinfo {volume} {97}},\ \bibinfo {pages} {165430} (\bibinfo {year} {2018})}\BibitemShut {NoStop}%
\bibitem [{\citenamefont {Giovannetti}\ \emph {et~al.}(2015{\natexlab{b}})\citenamefont {Giovannetti}, \citenamefont {Capone}, \citenamefont {van~den Brink},\ and\ \citenamefont {Ortix}}]{kekuletexturesgianluca2015}%
  \BibitemOpen
  \bibfield  {author} {\bibinfo {author} {\bibfnamefont {G.}~\bibnamefont {Giovannetti}}, \bibinfo {author} {\bibfnamefont {M.}~\bibnamefont {Capone}}, \bibinfo {author} {\bibfnamefont {J.}~\bibnamefont {van~den Brink}},\ and\ \bibinfo {author} {\bibfnamefont {C.}~\bibnamefont {Ortix}},\ }\bibfield  {title} {\bibinfo {title} {Kekul\'e textures, pseudospin-one dirac cones, and quadratic band crossings in a graphene-hexagonal indium chalcogenide bilayer},\ }\href {https://doi.org/10.1103/PhysRevB.91.121417} {\bibfield  {journal} {\bibinfo  {journal} {Phys. Rev. B}\ }\textbf {\bibinfo {volume} {91}},\ \bibinfo {pages} {121417} (\bibinfo {year} {2015}{\natexlab{b}})}\BibitemShut {NoStop}%
\bibitem [{\citenamefont {Garcia}\ \emph {et~al.}(2011)\citenamefont {Garcia}, \citenamefont {de~Lima}, \citenamefont {Assali},\ and\ \citenamefont {Justo}}]{GarciaJoelsonGroupIVGraphene}%
  \BibitemOpen
  \bibfield  {author} {\bibinfo {author} {\bibfnamefont {J.~C.}\ \bibnamefont {Garcia}}, \bibinfo {author} {\bibfnamefont {D.~B.}\ \bibnamefont {de~Lima}}, \bibinfo {author} {\bibfnamefont {L.~V.~C.}\ \bibnamefont {Assali}},\ and\ \bibinfo {author} {\bibfnamefont {J.~F.}\ \bibnamefont {Justo}},\ }\bibfield  {title} {\bibinfo {title} {Group iv graphene- and graphane-like nanosheets},\ }\href {https://doi.org/10.1021/jp203657w} {\bibfield  {journal} {\bibinfo  {journal} {The Journal of Physical Chemistry C}\ }\textbf {\bibinfo {volume} {115}},\ \bibinfo {pages} {13242} (\bibinfo {year} {2011})}\BibitemShut {NoStop}%
\bibitem [{\citenamefont {Carvalho}\ \emph {et~al.}(2013)\citenamefont {Carvalho}, \citenamefont {Ribeiro},\ and\ \citenamefont {Castro~Neto}}]{CarvalhoBandNesting2013}%
  \BibitemOpen
  \bibfield  {author} {\bibinfo {author} {\bibfnamefont {A.}~\bibnamefont {Carvalho}}, \bibinfo {author} {\bibfnamefont {R.~M.}\ \bibnamefont {Ribeiro}},\ and\ \bibinfo {author} {\bibfnamefont {A.~H.}\ \bibnamefont {Castro~Neto}},\ }\bibfield  {title} {\bibinfo {title} {Band nesting and the optical response of two-dimensional semiconducting transition metal dichalcogenides},\ }\href {https://doi.org/10.1103/PhysRevB.88.115205} {\bibfield  {journal} {\bibinfo  {journal} {Phys. Rev. B}\ }\textbf {\bibinfo {volume} {88}},\ \bibinfo {pages} {115205} (\bibinfo {year} {2013})}\BibitemShut {NoStop}%
\bibitem [{\citenamefont {Mennel}\ \emph {et~al.}(2020)\citenamefont {Mennel}, \citenamefont {Smejkal}, \citenamefont {Linhart}, \citenamefont {Burgd\"{o}rfer}, \citenamefont {Libisch},\ and\ \citenamefont {Mueller}}]{mennel2020band}%
  \BibitemOpen
  \bibfield  {author} {\bibinfo {author} {\bibfnamefont {L.}~\bibnamefont {Mennel}}, \bibinfo {author} {\bibfnamefont {V.}~\bibnamefont {Smejkal}}, \bibinfo {author} {\bibfnamefont {L.}~\bibnamefont {Linhart}}, \bibinfo {author} {\bibfnamefont {J.}~\bibnamefont {Burgd\"{o}rfer}}, \bibinfo {author} {\bibfnamefont {F.}~\bibnamefont {Libisch}},\ and\ \bibinfo {author} {\bibfnamefont {T.}~\bibnamefont {Mueller}},\ }\bibfield  {title} {\bibinfo {title} {Band nesting in two-dimensional crystals: An exceptionally sensitive probe of strain},\ }\href {https://doi.org/10.1021/acs.nanolett.0c00694} {\bibfield  {journal} {\bibinfo  {journal} {Nano letters}\ }\textbf {\bibinfo {volume} {20}},\ \bibinfo {pages} {4242} (\bibinfo {year} {2020})}\BibitemShut {NoStop}%
\bibitem [{\citenamefont {Mak}\ \emph {et~al.}(2008)\citenamefont {Mak}, \citenamefont {Sfeir}, \citenamefont {Wu}, \citenamefont {Lui}, \citenamefont {Misewich},\ and\ \citenamefont {Heinz}}]{MakMeasurementOpticalGraphene2008}%
  \BibitemOpen
  \bibfield  {author} {\bibinfo {author} {\bibfnamefont {K.~F.}\ \bibnamefont {Mak}}, \bibinfo {author} {\bibfnamefont {M.~Y.}\ \bibnamefont {Sfeir}}, \bibinfo {author} {\bibfnamefont {Y.}~\bibnamefont {Wu}}, \bibinfo {author} {\bibfnamefont {C.~H.}\ \bibnamefont {Lui}}, \bibinfo {author} {\bibfnamefont {J.~A.}\ \bibnamefont {Misewich}},\ and\ \bibinfo {author} {\bibfnamefont {T.~F.}\ \bibnamefont {Heinz}},\ }\bibfield  {title} {\bibinfo {title} {Measurement of the optical conductivity of graphene},\ }\href {https://doi.org/10.1103/PhysRevLett.101.196405} {\bibfield  {journal} {\bibinfo  {journal} {Phys. Rev. Lett.}\ }\textbf {\bibinfo {volume} {101}},\ \bibinfo {pages} {196405} (\bibinfo {year} {2008})}\BibitemShut {NoStop}%
\bibitem [{\citenamefont {Falkovsky}(2008)}]{LAFalkovsky2008}%
  \BibitemOpen
  \bibfield  {author} {\bibinfo {author} {\bibfnamefont {L.~A.}\ \bibnamefont {Falkovsky}},\ }\bibfield  {title} {\bibinfo {title} {Optical properties of graphene},\ }\href {https://doi.org/10.1088/1742-6596/129/1/012004} {\bibfield  {journal} {\bibinfo  {journal} {Journal of Physics: Conference Series}\ }\textbf {\bibinfo {volume} {129}},\ \bibinfo {pages} {012004} (\bibinfo {year} {2008})}\BibitemShut {NoStop}%
\bibitem [{\citenamefont {Horng}\ \emph {et~al.}(2011)\citenamefont {Horng}, \citenamefont {Chen}, \citenamefont {Geng}, \citenamefont {Girit}, \citenamefont {Zhang}, \citenamefont {Hao}, \citenamefont {Bechtel}, \citenamefont {Martin}, \citenamefont {Zettl}, \citenamefont {Crommie}, \citenamefont {Shen},\ and\ \citenamefont {Wang}}]{HorngDrudeCondGraphene2011}%
  \BibitemOpen
  \bibfield  {author} {\bibinfo {author} {\bibfnamefont {J.}~\bibnamefont {Horng}}, \bibinfo {author} {\bibfnamefont {C.-F.}\ \bibnamefont {Chen}}, \bibinfo {author} {\bibfnamefont {B.}~\bibnamefont {Geng}}, \bibinfo {author} {\bibfnamefont {C.}~\bibnamefont {Girit}}, \bibinfo {author} {\bibfnamefont {Y.}~\bibnamefont {Zhang}}, \bibinfo {author} {\bibfnamefont {Z.}~\bibnamefont {Hao}}, \bibinfo {author} {\bibfnamefont {H.~A.}\ \bibnamefont {Bechtel}}, \bibinfo {author} {\bibfnamefont {M.}~\bibnamefont {Martin}}, \bibinfo {author} {\bibfnamefont {A.}~\bibnamefont {Zettl}}, \bibinfo {author} {\bibfnamefont {M.~F.}\ \bibnamefont {Crommie}}, \bibinfo {author} {\bibfnamefont {Y.~R.}\ \bibnamefont {Shen}},\ and\ \bibinfo {author} {\bibfnamefont {F.}~\bibnamefont {Wang}},\ }\bibfield  {title} {\bibinfo {title} {Drude conductivity of dirac fermions in graphene},\ }\href {https://doi.org/10.1103/PhysRevB.83.165113} {\bibfield  {journal} {\bibinfo  {journal} {Phys. Rev. B}\ }\textbf {\bibinfo {volume} {83}},\ \bibinfo
  {pages} {165113} (\bibinfo {year} {2011})}\BibitemShut {NoStop}%
\bibitem [{\citenamefont {D\'ora}\ \emph {et~al.}(2011)\citenamefont {D\'ora}, \citenamefont {Kailasvuori},\ and\ \citenamefont {Moessner}}]{DoraLatticeGeneralization2011}%
  \BibitemOpen
  \bibfield  {author} {\bibinfo {author} {\bibfnamefont {B.}~\bibnamefont {D\'ora}}, \bibinfo {author} {\bibfnamefont {J.}~\bibnamefont {Kailasvuori}},\ and\ \bibinfo {author} {\bibfnamefont {R.}~\bibnamefont {Moessner}},\ }\bibfield  {title} {\bibinfo {title} {Lattice generalization of the dirac equation to general spin and the role of the flat band},\ }\href {https://doi.org/10.1103/PhysRevB.84.195422} {\bibfield  {journal} {\bibinfo  {journal} {Phys. Rev. B}\ }\textbf {\bibinfo {volume} {84}},\ \bibinfo {pages} {195422} (\bibinfo {year} {2011})}\BibitemShut {NoStop}%
\bibitem [{\citenamefont {Koshino}\ and\ \citenamefont {Son}(2019)}]{MoirephononsKoshino2019}%
  \BibitemOpen
  \bibfield  {author} {\bibinfo {author} {\bibfnamefont {M.}~\bibnamefont {Koshino}}\ and\ \bibinfo {author} {\bibfnamefont {Y.-W.}\ \bibnamefont {Son}},\ }\bibfield  {title} {\bibinfo {title} {Moir\'e phonons in twisted bilayer graphene},\ }\href {https://doi.org/10.1103/PhysRevB.100.075416} {\bibfield  {journal} {\bibinfo  {journal} {Phys. Rev. B}\ }\textbf {\bibinfo {volume} {100}},\ \bibinfo {pages} {075416} (\bibinfo {year} {2019})}\BibitemShut {NoStop}%
\bibitem [{\citenamefont {Ochoa}\ and\ \citenamefont {Asenjo-Garcia}(2020)}]{OchoaFlatBandsMoire2020}%
  \BibitemOpen
  \bibfield  {author} {\bibinfo {author} {\bibfnamefont {H.}~\bibnamefont {Ochoa}}\ and\ \bibinfo {author} {\bibfnamefont {A.}~\bibnamefont {Asenjo-Garcia}},\ }\bibfield  {title} {\bibinfo {title} {Flat bands and chiral optical response of moir\'e insulators},\ }\href {https://doi.org/10.1103/PhysRevLett.125.037402} {\bibfield  {journal} {\bibinfo  {journal} {Phys. Rev. Lett.}\ }\textbf {\bibinfo {volume} {125}},\ \bibinfo {pages} {037402} (\bibinfo {year} {2020})}\BibitemShut {NoStop}%
\end{thebibliography}%
\end{document}